\documentclass[sigconf, authorversion, noacm]{acmart}

\usepackage{microtype}
\usepackage{eucal}
\usepackage{xcolor}
\usepackage{amsmath}
\usepackage{bm}
\usepackage{wrapfig}

\usepackage{cleveref}
\usepackage{wrapfig}
\usepackage{subfig}
\usepackage{multirow} 
\usepackage{csvsimple}
\usepackage{mathtools}
\usepackage{mathrsfs}
\usepackage{listings}
\usepackage{booktabs}

\usepackage{listings}
\usepackage{xcolor}

\definecolor{codegreen}{rgb}{.7, .7, .7}
\definecolor{codeblue}{rgb}{.0, .0, .82}
\definecolor{backcolour}{rgb}{.96, .96, .96}

\lstdefinestyle{mystyle}{
    backgroundcolor=\color{backcolour},   
    commentstyle=\color{codegreen},
    keywordstyle=\color{codeblue},
    basicstyle=\ttfamily\footnotesize,
    breakatwhitespace=false,
    breaklines=true,
    captionpos=b,
    keepspaces=true,
    numbers=left,
    numbersep=5pt,
    showspaces=false,
    showstringspaces=false,
    showtabs=false,                  
    tabsize=2
}
\usepackage[linesnumbered, ruled]{algorithm2e}
\SetKwInput{Kw}{input}
\SetKwComment{Comment}{$\vartriangleright$\ }{}

\SetCommentSty{mycommfont}
\SetKwProg{Struct}{struct}{}{}

\usepackage{tabularx}

\newcolumntype{Y}{>{\centering\arraybackslash}X}
\newcolumntype{L}{>{\arraybackslash}X}

\crefname{figure}{fig.}{figures}
\Crefname{figure}{Fig.}{Figures}
\crefname{table}{tab.}{tables}
\Crefname{table}{Table}{Tables}
\crefname{algorithm}{alg.}{algorithms}
\Crefname{algorithm}{Algorithm}{Algorithms}
\crefname{appendix}{app.}{appendices}
\Crefname{appendix}{App.}{Appendices}

\newcommand{\bl}[1]{{\color{purple}}}

\newcommand{\abs}[1]{\lvert #1 \rvert}
\newcommand{\norm}[1]{\lvert\lvert #1 \rvert\rvert}

\newcommand{\site}{\ensuremath{\mathbf{p}}}

\newcommand{\object}{\ensuremath{o}}

\newcommand{\weight}{\ensuremath{\psi}}
\newcommand{\weightset}{\ensuremath{\mathbf{\Psi}}}

\newcommand{\powercell}{V}
\newcommand{\powerbisector}{V}

\newcommand{\point}{\mathbf{x}}
\newcommand{\prescribedvolume}{\nu}
\newcommand{\totalvolume}{V}

\graphicspath{%
    {./figures/}%
}

\AtBeginDocument{%
  }

\setcopyright{acmlicensed}
\copyrightyear{2018}
\acmYear{2018}
\acmDOI{XXXXXXX.XXXXXXX}

\begin{document}

\title{More Power to the Particles: Analytic Geometry for Partial Optimal Transport-based Fluid Simulation}

\author{Cyprien Plateau--Holleville}
\orcid{0000-0003-1510-557X}
\email{cyprien.plateau-holleville@inria.fr}
\affiliation{%
  \institution{Inria Paris-Saclay}
  \city{Paris-Saclay}
  \country{France}
}

\author{Bruno Lévy}
\orcid{0000-0002-7007-3219}
\email{Bruno.Levy@inria.fr}
\affiliation{%
    \institution{Inria Paris-Saclay, Université Paris-Saclay, CNRS, Laboratoire de Mathématiques d'Orsay}
    \country{France}
}

\renewcommand{\shortauthors}{Plateau--Holleville et al.}

\begin{CCSXML}
<ccs2012>
<concept>
<concept_id>10010147.10010371.10010352.10010379</concept_id>
<concept_desc>Computing methodologies~Physical simulation</concept_desc>
<concept_significance>500</concept_significance>
</concept>
<concept>
<concept_id>10003752.10010061.10010063</concept_id>
<concept_desc>Theory of computation~Computational geometry</concept_desc>
<concept_significance>300</concept_significance>
</concept>
<concept>
<concept_id>10010147.10010169.10010170.10010174</concept_id>
<concept_desc>Computing methodologies~Massively parallel algorithms</concept_desc>
<concept_significance>300</concept_significance>
</concept>
</ccs2012>
\end{CCSXML}

\ccsdesc[500]{Computing methodologies~Physical simulation}
\ccsdesc[300]{Theory of computation~Computational geometry}
\ccsdesc[300]{Computing methodologies~Massively parallel algorithms}

\keywords{Fluid Simulation, Semi-Discrete Optimal Transport, Laguerre Diagram, GPGPU}

\begin{teaserfigure}
    \includegraphics[width=\textwidth]{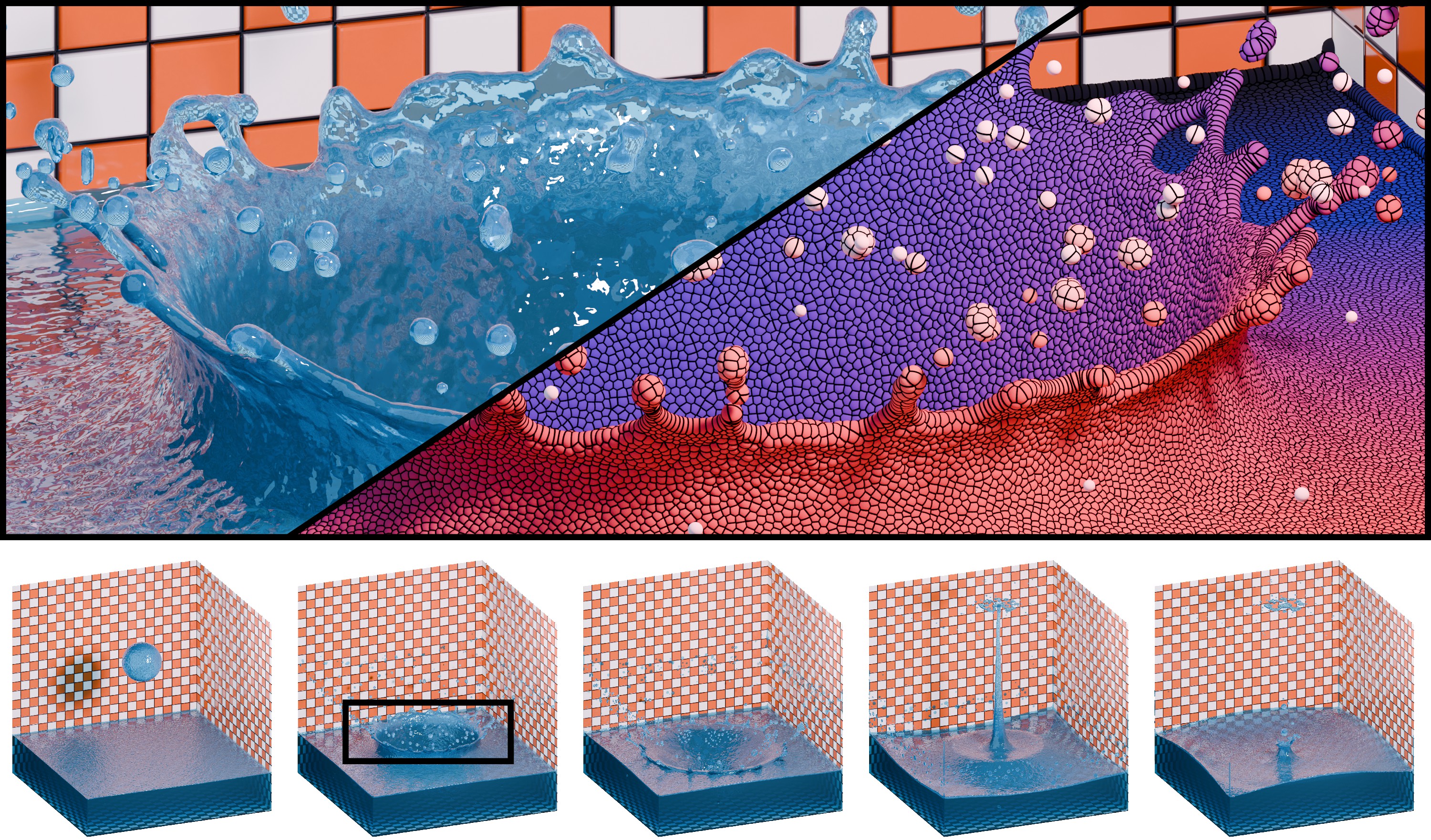}  
    \caption{Large-scale fluid simulation (left) computed with our analytic partial optimal transport solver with 2 million cells (right). Thin droplets of different sizes that split and merge are accurately tracked across the simulation while accurately preserving volume. The fluid is represented with cells characterized by the intersection between balls and Laguerre cells, that we analytically evaluate. Thanks to this representation, the surface of the liquid can be easily reconstructed as a mesh to be used in standard rendering pipelines or directly displayed with the custom rendering engine described here.}
    \label{fig:teaser}
\end{teaserfigure}

\begin{abstract}
We propose unified data structures and algorithms for free-surface fluid simulations based on partial optimal transport, such as the Power Particles method or Gallouët-Mérigot's scheme. Such methods previously relied on a discretization of the cells by leveraging a classical convex cell clipping algorithm. However, this results in a heavy computational cost and a coarse approximation of the evaluated quantities. In contrast, we propose to analytically construct the generalized Laguerre cells characterized by intersections between Laguerre cells and spheres. This makes it possible to accurately compute the differential quantities used by the Newton algorithm, that is, the areas of the (curved) facets and the volumes of the (generalized) Laguerre cells. This significantly improves the convergence of the Newton algorithm, hence the robustness of the simulations, even in challenging scenarios with high velocities and chocs. Moreover, this drastically reduces the computational cost as compared to previous works. Based on our data structure, we propose a framework that combines (1) the numerical solution mechanism for partial optimal transport, (2) the fluid simulation scheme and (3) the rendering. The aforementioned three components are implemented on the GPU, providing further speedup and avoiding data transfers. This is made possible by the compactness of our data structure combined with a massively parallel implementation. We report the result of numerical experiments featuring highly detailed, large-scale simulations and high variations of physical properties within the same simulation.
\end{abstract}

\maketitle

\section{Introduction}

Simulating complex fluid behaviors with a computer still represents a challenge. Notably, they are difficult to implement with strong guarantees regarding the properties of the simulation. This is especially true with volume conservation, which characterizes incompressible liquids. The chosen representation of the fluid in the computer has a direct impact on the support of topological changes (\textit{e.g.} with droplets that split and merge) or geometric interactions between the fluid, the domain boundary or other objects. Additionally, the resolution of the discretization strongly constrains the ability of the scheme to capture detailed effects like the formation of small droplets.

Multiple representations have been presented based on either an Eulerian representation, with fields attached to a 3D grid (\textit{e.g.} LBM~\cite{McNamara1988}), a Lagrangian representation, with particles that follow fluid motion (\textit{e.g.} SPH~\cite{Gingold1977}), or a hybridization of both (\textit{e.g.} Stable Fluids~\cite{Stam1999}). Despite the tremendous progress made towards robust and efficient simulations, it is still difficult to track very thin interfaces or to represent fine sheets of fluids while conserving volume and tracking the free surface accurately. 
 
A whole family of recent works in the Lagrangian setting were developed relying on a special case of a semi-discrete optimal transport problem that prescribes a fixed fluid quantity per particle~\cite{Goes2015, Gallouet2017, Levy2022, Qu2022}. Thanks to their theoretical background, these methods offer strong guarantees on volume preservation which is enforced at every step of the simulation, in contrast with previous Lagrangian discretizations that constrain it indirectly.

Despite their advantages, semi-discrete optimal transport-based solvers face severe constraints linked to their geometric representation, defined by a Laguerre diagram~\cite{Aurenhammer1987}. Laguerre cells, describing the Lagrangian mesh discretization, are unbounded and cannot be directly used for free surface simulations. Previous works handled these limitations by either adding ghost cells to the representation~\cite{Goes2015}, artificially limiting liquid cells, or by relying on discrete kernels and regularized optimal transport solvers~\cite{Qu2022}, lowering accuracy. In contrast, another line of work introduced a formulation of this issue as a partial optimal transport problem~\cite{Levy2022}, providing an idealized mathematical definition, but implemented in practice through a polygonal discretization of the free surface.

Motivated by the elegant formulation provided by existing partial optimal transport-based solvers and their accurate results, we address their main limitation by introducing a novel framework targeting an exact representation of the free surface without explicit discretization. Compared to previous solvers, our method is fast, without compromise between quality and performance, in the sense that a precise convergence of the optimal transport is obtained. More specifically, this is made possible by the following contributions:\begin{itemize}
    \item we propose a new \emph{analytic} computation of the differential quantities involved in semi-discrete partial optimal transport (volumes and areas of generalized Laguerre cells and facets) which is robust, precise and fast to compute;
    \item deriving the geometric characterization, we \emph{efficiently} build a \emph{compact representation} used by the Newton solver. In addition, it naturally defines an accurate representation of the fluid and its surface;
    \item thanks to this compact data structure, we provide \emph{a massively parallel implementation running on the GPU}. This includes matrix assembly, linear solver and the Newton optimization algorithm;
    \item leveraging this representation, we provide a \emph{rendering engine} directly relying on the underlying data structure.
\end{itemize} Our fluid dynamics solver is able to simulate complex effects, including surface tension and viscosity, while handling large parameters ranges, even in the same simulation, that can for instance combine very viscous fluids with non-viscous ones. Our implementation will be made available after publication.

\section{Related works}
\label{ss:related-works}

Fluid solvers can be classified based on their underlying representation, which can either represent fields (\emph{Eulerian}) or follow the movement of the fluid (\emph{Lagrangian}). While an Eulerian representation supports the implementation of an accurate solver, it involves a spatial discretization defined over the complete domain, describing local fluid quantities at fixed locations. In contrast, the parameters of a Lagrangian representation are following the motion of the fluid, providing more detailed effects but traditionally offering fewer guarantees regarding the incompressibility. Based on these characteristics, numerous works have proposed alternatives by developing new discretization. Recently introduced, optimal transport-based solvers~\cite{Benamou2000, Goes2015, Gallouet2017} are an interesting family of Lagrangian methods with strong guarantees regarding mass preservation and addressing some limitations of previous representations. Starting from historical and standard solvers based on classical discretization (\Cref{ss:eulerian-lagrangian}), we then present the most relevant previous works with a focus on optimal transport-based methods (\Cref{ss:ot}).

\subsection{Eulerian and Lagrangian}
\label{ss:eulerian-lagrangian}

Classical Eulerian and Lagrangian fluid representations have been extensively developed in Computed Graphics and Computational Fluid Dynamics but they still suffer from limitations regarding accuracy, matter tracking and ease of implementation. 

\begin{figure}[b]
    \subfloat[Smoothed particles\label{fig:connectivity-field}]{\includegraphics[width=.3\columnwidth]{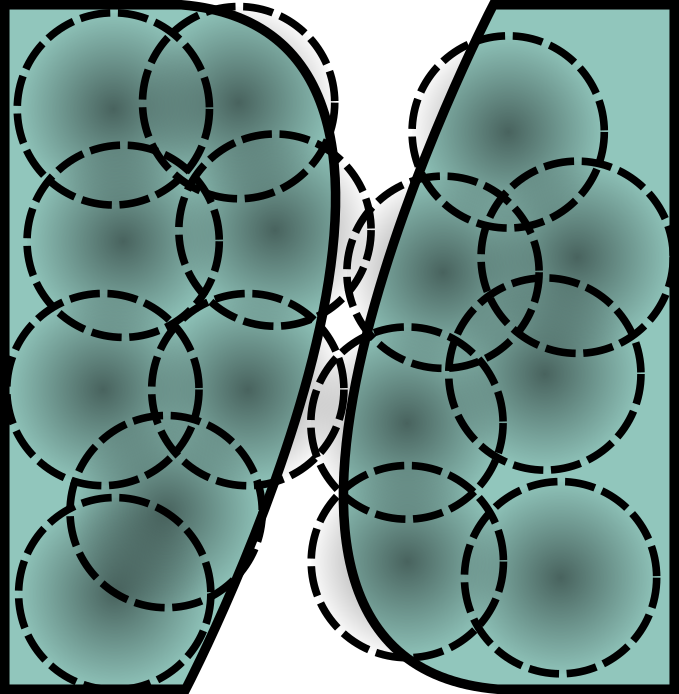}}\hfill
    \subfloat[Mesh\label{fig:connectivity-mesh}]{\includegraphics[width=.3\columnwidth]{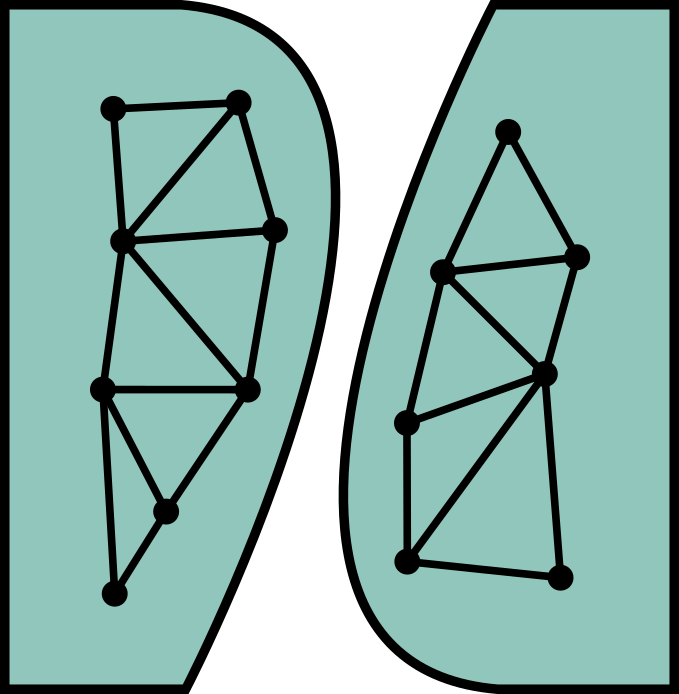}}\hfill
    \subfloat[Laguerre\label{fig:connectivity-voronoi}]{\includegraphics[width=.3\columnwidth]{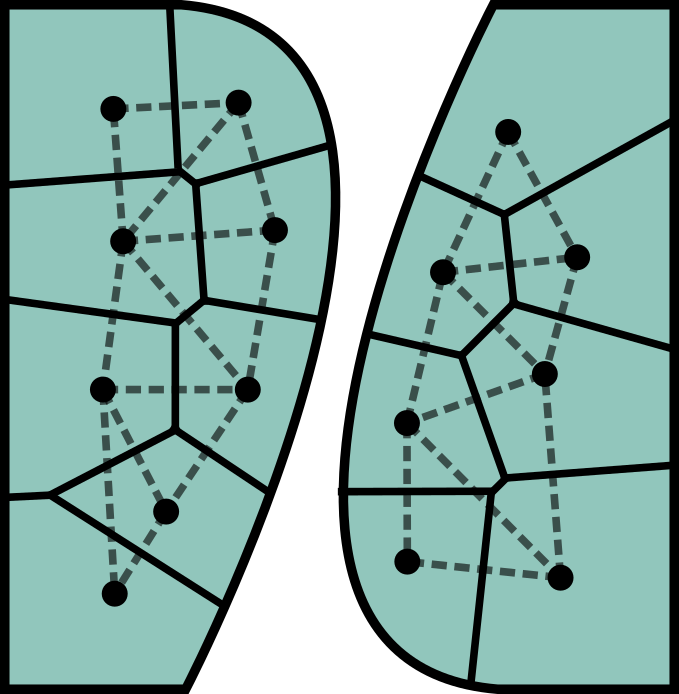}}
    \caption{Illustration of two sheets of fluids discretized using different Lagrangian representations and the corresponding fluid elements neighborhoods. (a) It may be difficult to determine a well behaved surface from smoothed particles in ambiguous configuration (intersecting interaction rings while parts of different fluid sheets). (b) In contrast, mesh-based representations accurately track the neighborhood, but may need topology updates across the simulation. (c) Laguerre-based representation defines the neighborhoods implicitly.}
    \label{fig:connectivity}
\end{figure}

\paragraph{Eulerian Fluid Representation} Historically, a Eulerian setting has been highly used in hybrid solvers coupled with a Lagrangian representation~\cite{Brackbill1986, Stam1999, Stomakhin2013, Jiang2015}, but is often limited by complex implementations relying on heuristics. The \emph{Lattice-Boltzmann method} (LBM)~\cite{McNamara1988} is able to robustly simulate highly complex behaviors~\cite{Li2023, Li2024, Wang2025}, in both Computer Graphics~\cite{Guo2017} and Computational Fluid Dynamics~\cite{Lehmann2022} communities. Despite its advantages, it often faces severe limitations, such as considerable memory consumption and possible artifacts and aliasing linked to the resolution of the Euler grid.

\paragraph{Lagrangian Fluid Representation} \emph{Smoothed Particles Hydrodynamics} (SPH)~\cite{Gingold1977} is a popular numerical scheme in the Computer Graphics community (one may refer to \cite{Koschier2022} for an introduction to the topic) based on the approximation of the density field with discretization points. This representation offers multiple advantages, notably regarding the trade-off between computational requirements and amount of details. However, it is well known that producing incompressible flows through divergence-free velocity is difficult to guarantee and requires specific adaptations~\cite{Bender2017}.

\paragraph{Modeling Complex Effects} Discretizing complex fluid behaviors remains a challenge linked to the selected representation. This is often due to the discretization of differential operators, not always satisfying the properties of their continuous counterpart. Numerous previous works have then proposed various ways to overcome this issue by specific expressions of the operators for effects like viscosity and surface tension. Viscosity is a complex phenomenon \emph{that tends to homogenize the velocity field}. Modeled as a force that is proportional to the Laplacian of the velocity field, it strongly suffers from inexact incompressible velocity. Various Eulerian or hybrid methods~\cite{Stam1999} have then been presented to precisely simulate this effect, but handling high viscosity using Lagrangian methods remains difficult~\cite{Weiler2018}. Surface tension is a force caused by the \emph{asymmetry of attraction} between liquid elements near its interface. However, the limits of the fluid are often difficult to estimate precisely.  This has motivated methods specially tailored for explicit handling of \emph{surface effects} like surface area minimization~\cite{Akinci2013, He2014, Huber15, Da2016, Xing2022} through additional processing, or implicit integration~\cite{Jeske2023}.

\subsection{Optimal Transport-based Simulation}
\label{ss:ot}

Previous developments have shown that one of the most difficult aspects of fluid modelization remains the accurate conservation of physical quantities and adaptive spatial discretization. Following this observation, it seems advantageous to define models that enforce these criteria \emph{by construction}. Starting from a Lagrangian mesh to preserve moving coordinates while ensuring well-defined connectivity (\Cref{fig:connectivity}), a numerical approach must be defined to constrain a prescribed fluid volume to each fluid element throughout the simulation. This can be achieved by solving a special case of an optimal transport problem. Once defined, this representation must then be adapted to the more general setting with free surfaces, to accurately track its interfaces. 

\paragraph{Lagrangian Mesh Representation} A Lagrangian mesh representation describes the fluid as a set of cells $(\powercell_i)_{i=1}^{n}$, each of them depending on some parameters $\pi_i(t)$. The evolution of the parameters with respect to time fully determines the evolution of the cells and then of the fluid. The cells are linked by explicit connections which can be defined or deduced from their parameters, in contrast with Lagrangian meshless representation like SPH. However, fixed connections suffer from strong topological changes that break the interaction between fluid cells. For instance, one can use evolving tetrahedral meshes as a representation, the parameters of which are then the coordinates at the vertices. In such a case, an important deformation would entangle the mesh. Since this impacts the validity of the simulation, it often requires remeshing across the simulation~\cite{Wicke2010, Clausen2013}. To address this issue, hybrid methods have been proposed, such as the Arbitrary Lagrangian Euler (ALE) \cite{Hirt1974}, but may still fail to accurately enforce volume conservation.

\paragraph{Volume-Constrained Lagrangian Mesh} Our goal is now to define a Lagrangian mesh representation, where each cell $\powercell_i$ is parameterized by the couple $\pi_i \coloneqq \{\site_i, \prescribedvolume_i\} \in \mathbb R^d \times \mathbb R^+$, where $\site_i$ is a point of the simulation domain, and $\prescribedvolume_i$ is a prescribed volume for the cell. Clearly, we have $\sum_i^n \prescribedvolume_i = \totalvolume$, with $\totalvolume$ the total fluid volume. We chose a Laguerre diagram (also referred to as Power diagram) as the representation of our cells, defined by \[
    \powercell_i = \left\{ x \, \mid \, \norm{\point - \site_i}^2 - \weight_i \leq \norm{\point - \site_j}^2 - \weight_j\right\}.
\] As can be seen, the Laguerre diagram is parameterized by the set of points $(\site_i)_{i=1}^n$ and a vector $\weightset = (\weight_i)_{i=1}^n \in \mathbb R^n$ of parameters, called the \emph{weights} of the Laguerre diagram. It then remains to determine the weight vector $\weightset$ from our parameters $\left(\pi_i \coloneqq \{\site_i, \prescribedvolume_i\}\right)_{i=1}^N$. It is well known in semi-discrete transport theory~\cite{Brenier1991, Aurenhammer1992, Kitagawa2019} that for a given set of parameters $(\pi_i)_{i=1}^N$, there exists a unique weight vector $\weightset$ (up to a constant) such that the value prescriptions are satisfied (\textit{i.e.} $|\powercell_i(\weightset)| = \prescribedvolume_i \; \forall i$). This vector $\weightset$ is obtained as the solution of a convex optimization problem, maximizing the objective function: \[
    K\left(\weightset\right) = \sum_i{\int_{\powercell_i(\weightset)}[\norm{\point - \site_i}^2 - \weight_i]d\point} + \sum_i{\psi_i\prescribedvolume_i},
\] where $\weightset$ is called the weight vector. Hence, maximizing $K$, through a Newton optimization algorithm (described in Algorithm~\ref{alg:volume-optimization}), gives the unique weight vector (up to a translation) that satisfies the volume constrains~\cite{Aurenhammer1992, Kitagawa2019}. The set of cells $\powercell_i$ is then directly derived from the set of parameters $(\pi_i)_{i=1}^N$. Interestingly, the connectivity of the mesh implicitly emerges from the solution of the optimization problem.

\begin{algorithm}[t]
    \DontPrintSemicolon

	\SetArgSty{upshape}
    \KwData{The sites $(\site_i)_{i=1}^n$ and prescribed volumes $(\prescribedvolume_i)_{i=1}^n$}
    \KwResult{The unique, up to a translation, vector of weights $\weightset = (\weight_i)_{i=1}^n$ maximizing $K$}

	\caption{Laguerre Cells Volume Optimization}\label{alg:volume-optimization}
    
	$\weightset \leftarrow 0$\;
    \Comment{\textbf{Continue until error is below $\epsilon$}}
	\While{$\sup{(|\;|\powercell_i(\weightset)| - \prescribedvolume_i|)} > \epsilon$}{
    	Solve for $\mathbf{u}$ in $(\nabla^2K)\mathbf{u} = -\nabla \mathbf{u}$\;
        \Comment{\textbf{Compute $\alpha$ with KMT steps}}
        \While{$\inf{(|\powercell_i(\weight_i + \alpha \mathbf{u}_i)|)} > \tfrac{1}{2}\inf(\inf(\prescribedvolume_i), \inf(|\powercell_i(0)|))$}{
            $\alpha \leftarrow \tfrac{1}{2}\alpha$\;
    	}
        $\weightset \leftarrow \weightset + \alpha \mathbf{u}$ \;
	}
\end{algorithm}

\paragraph{Optimal Transport-based Computational Fluid Dynamics Representation} Following \citeauthor{Brenier1989}'s intuition on the relations between optimal transport, the least action principle and fluids~\cite{Brenier1989, Benamou2000}, multiple works have been proposed to effectively handle the fluid discretization through a volume-constrained Lagrangian mesh. As such, the Power Particles method has enabled impressive large-scale simulations of incompressible fluids~\cite{Goes2015}. Their method relies on the systematic correction of sites' center to their barycenter derived from a dedicated pressure representation. Alternatively, it is possible to approximately enforce incompressibility through a harmonic oscillator, used to smoothly project the motion onto the manifold of volume-preserving motions in a way that provably converges to the exact motion in the case of incompressible Euler fluids~\cite{Gallouet2017}. As studied by \citeauthor{Duque2023}, this describes a more natural definition of the fluid motion without the need of an explicit extraction of an incompressible velocity field~\cite{Duque2023}. From the strong guarantees given by optimal transport-based simulations, several works have proposed various additional developments. Notably, better performance can be achieved from an adaptive framework implemented on a GPU~\cite{Zhai2020}. To limit the computational requirements of the transportation plan, it is also possible to rely on a reformulation as a partition of unity~\cite{Qu2022} allowing the use of entropic regularization through the Sinkhorn algorithm~\cite{Cuturi13}. This significantly improves the performance as compared to \citeauthor{Goes2015} and is compatible with the definition of hybrid models for both fluid~\cite{Qu2022} and inelastic materials~\cite{Qu2023}. However, it results in a less accurate transport and introduces the use of an explicit Eulerian discretization of the domain for the resolution of the discrete optimal transport problem. 

\begin{figure}[b]
    \subfloat[Free surface\label{fig:representation-free-surface-surface}]{\includegraphics[width=.325\columnwidth]{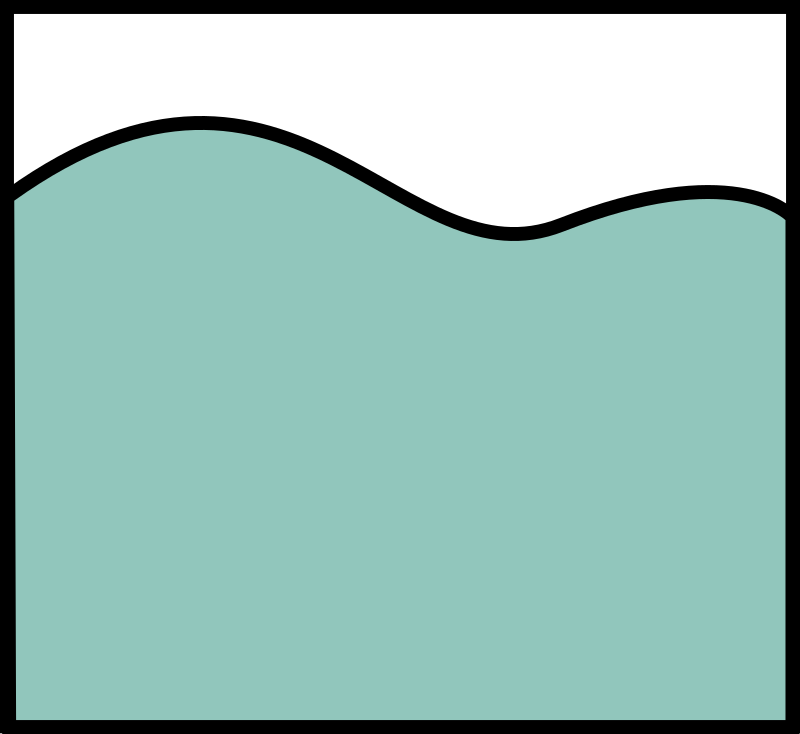}}\hfill
    \subfloat[Ghost Particles\label{fig:representation-free-surface-ghost-particle}]{\includegraphics[width=.325\columnwidth]{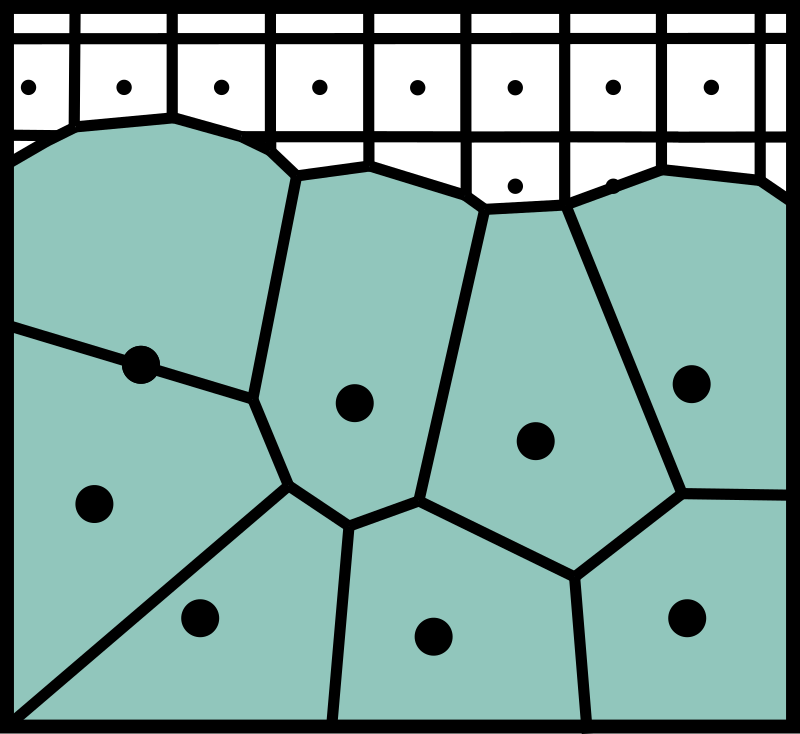}}\hfill
    \subfloat[Partial Transport\label{fig:representation-free-surface-partial}]{\includegraphics[width=.325\columnwidth]{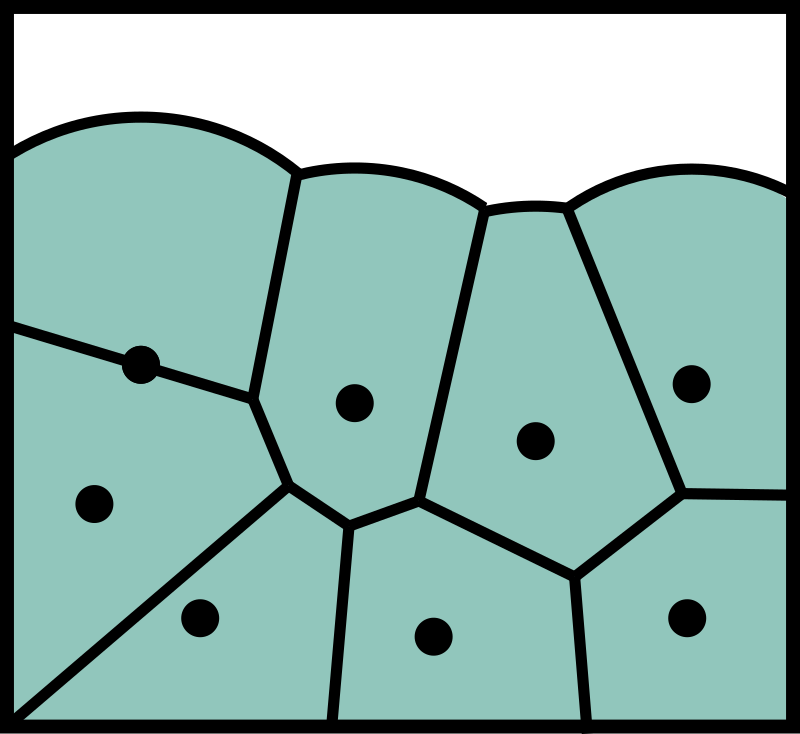}}
    \caption{Different ways of simulating (a) a fluid with a free surface using optimal transport-based fluid representation. (b) \citeauthor{Goes2015} use ghost particles distributed on a grid to bound Laguerre cells which involve a larger diagram (more costly to compute). (c) Partial optimal-transport characterizes the fluid boundary as Laguerre cells intersected with balls.}
    \label{fig:representation-free-surface}
\end{figure}

\begin{figure}[!b]
    \centering
    \subfloat[\cite{Goes2015, Levy2022}\label{fig:restricted-cell-discretized}]{\includegraphics[width=0.47\linewidth]{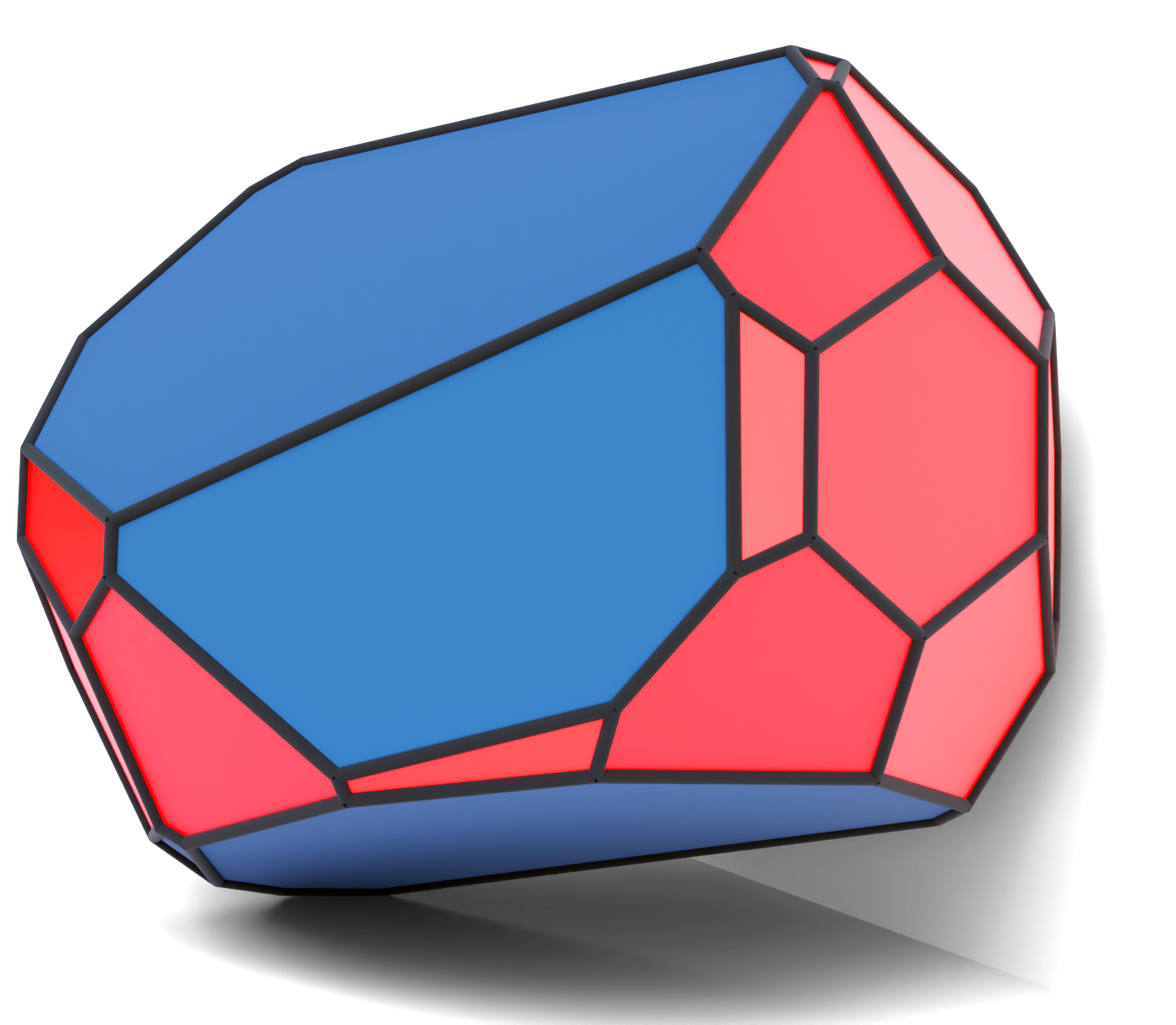}}
    \subfloat[Our restricted cell in $\mathbb R^3$\label{fig:restricted-cell}]{\includegraphics[width=0.47\linewidth]{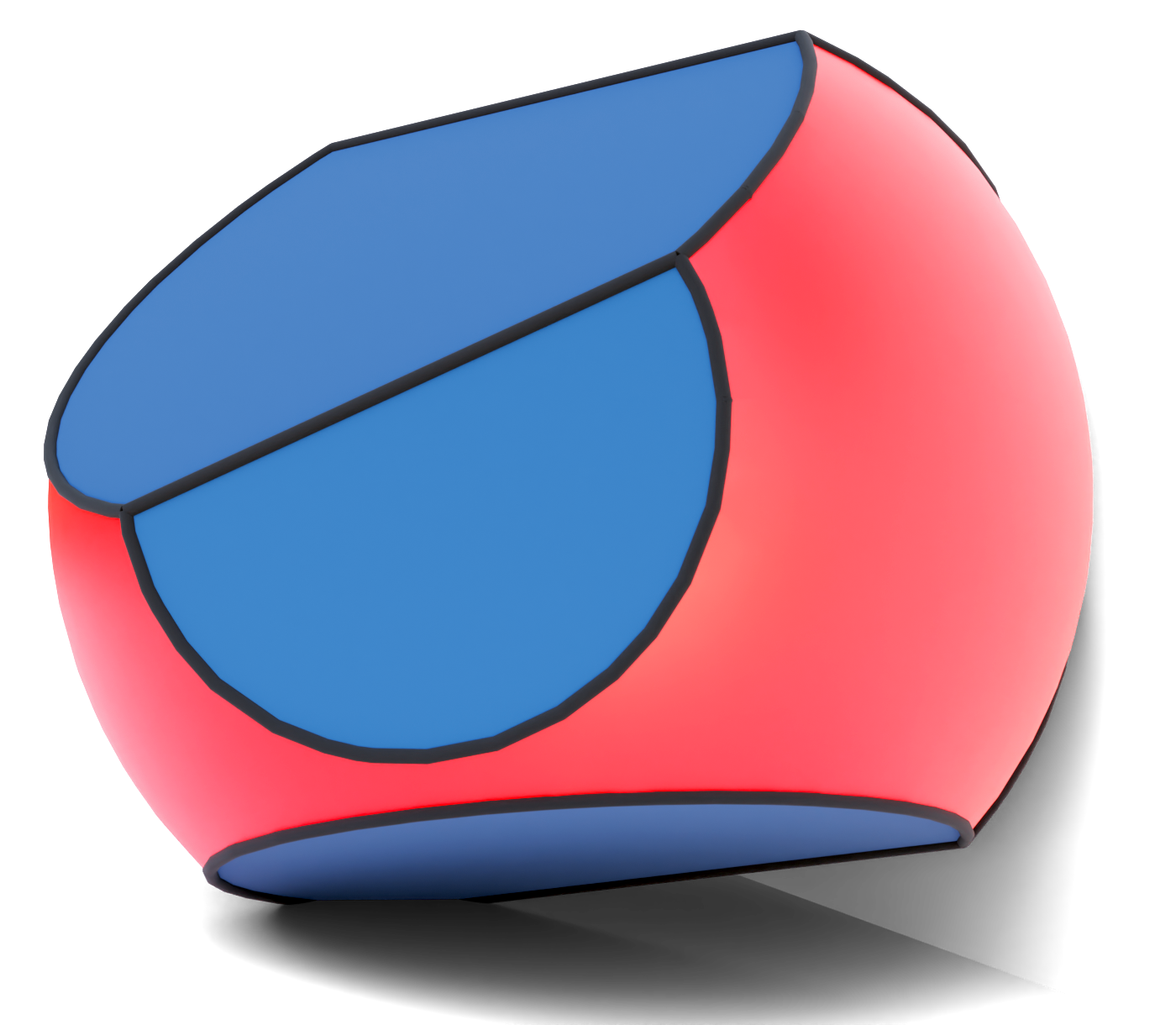}}\hfill
    \caption{Illustrations of the restriction geometry inherent to the free surface scheme. (a) Discretization as performed by previous methods~\cite{Goes2015, Levy2022}. (b) Restricted Laguerre facets (blue) and free surface (red). 
    }
    \label{fig:partial-geometry}
\end{figure}

\paragraph{Free Surface} The initial problem assigns a sub-volume of the whole domain to every cell. We now wish to represent a fluid that does not fill the entire simulation domain (\Cref{fig:representation-free-surface}). As the Laguerre diagram-based representation completely fills the domain, we must introduce a way to limit the cells such that they do not expend more than required. As proposed by a previous work~\cite{Goes2015}, this can be achieved by either explicitly representing ``air'' cells, or by bounding artificially the fluid cells with ghost cells. However, this requires tracking the interface of the fluid and results in a more costly computation of the Laguerre diagram (due to a large number of ``air'' cells). On the other hand, methods relying on a discrete setting can leverage the background grid~\cite{Qu2022}. To avoid this additional computational cost, it is possible to reformulate this problem so that the air volume is implicitly taken into account~\cite{Levy2022}, as also studied for crowd simulation~\cite{Leclerc2020}. Considering now that the number of ``air'' cells tends to infinity, then the set of ``air'' cells tends to an additional object $\object_0$ that fills the whole simulation domain and that receives the unassigned volume. In contrast with the initial formulation, this does not require an explicit discretization of the air volume while preserving exactly the prescribed volumes of the two phases. Again, this partial optimal transport problem can be solved by finding the unique weight vector $\weightset$. As a direct consequence of the invariability of the Laguerre Diagram to weight translation, we can fix the weight $\psi_0$ associated with $\object_0$ as $\psi_0 = 0$. The diagram of the resulting set of weighted sites $((\site_i, \psi_i))_{i=1}^n \cup (\object_0, \psi_0)$ is then composed of the original Laguerre cells $(\powercell_i)_{i = 1}^n$ modified by considering the cell $\powercell_0$. By noticing that $\norm{\object_0 - \point} = 0$ (because $\object_0$ fills up the entire domain), the cells are given by:\begin{equation}
    \powercell_i = \left\{ \point  \;\middle|\;
    \begin{aligned}
        & \norm{\point - \site_i}^2 - \psi_i \leq \norm{\point - \site_j}^2 - \psi_j, \; \forall j \neq i, \, j \neq 0, \\ 
        & \text{and} \; \norm{\point - \site_i}^2 - \psi_i \leq 0 \text{}
    \end{aligned}
    \right\}.
\end{equation} As illustrated in \Cref{fig:partial-geometry}, the so-defined cells corresponds to the intersections between the Laguerre cell $\powercell_i$ and the sphere $\Sigma_i$ centered on $\site_i$ and of radius $\sqrt{\psi_i}$. To avoid the difficulty of computing geometric quantities on spherical boundaries, previous works have relied on an explicit mesh discretization of the free surface~\cite{Levy2022} or restricting the generation of ``air particles'' to the vicinity of the fluid surface~\cite{Goes2015}. While their implementations can rely on existing convex polygon clipping and Laguerre diagrams, they also involve a memory and computational cost. More importantly, they may suffer from convergence issues, due to inaccurate computations of the quantities. We address this limitation and provide a dedicated framework based on this exact representation.

\subsection{Method Overview}
\label{ss:overview}

Given a set of cell representing a sub-volume of fluid, we wish to simulate and render an incompressible fluid motion. Multiple key components are then required: \begin{itemize}
    \item a \emph{partial optimal transport solver} relying on an analytic representation of the geometry (\cref{sec:analytic-geometry}) ensuring that every cell holds the correct fluid quantity;
    \item a \emph{GPU implementation} defined by efficient processing of our compact data structure (\cref{sec:gpu});
    \item a \emph{numeric simulation method for fluids}, which is classic (\cref{sec:physics});
    \item a \emph{dedicated ray-tracing engine} relying on the Laguerre diagram to display the fluid (\cref{sec:rendering}).
\end{itemize} Every element of our framework relies on the underlying modified Laguerre diagram. In contrast with previous works (\Cref{fig:restricted-cell-discretized}), this core feature is based on an analytic representation (\Cref{fig:restricted-cell}) of the geometry originating from the partial optimal transport problem. Thanks to this analytic representation, defined by the intersection between the Laguerre cell and the sphere described by its parameters, we are able to accurately track the fluid surface as well as the differential quantities involved in the solver and the numerical simulation.

\section{Partial Optimal Transport Solver}
\label{sec:analytic-geometry}

In this section, we describe the analytic computation of the geometry required to solve the semi-discrete partial optimal transport problem. We first recall the geometric components require to solve the optimal transport problem, provide a decomposition compatible with the required quantity, and then a method to compute it efficiently.

\begin{figure}[!t]
    \centering
    \subfloat[Pyramid decomposition]{\includegraphics[width=.47\columnwidth]{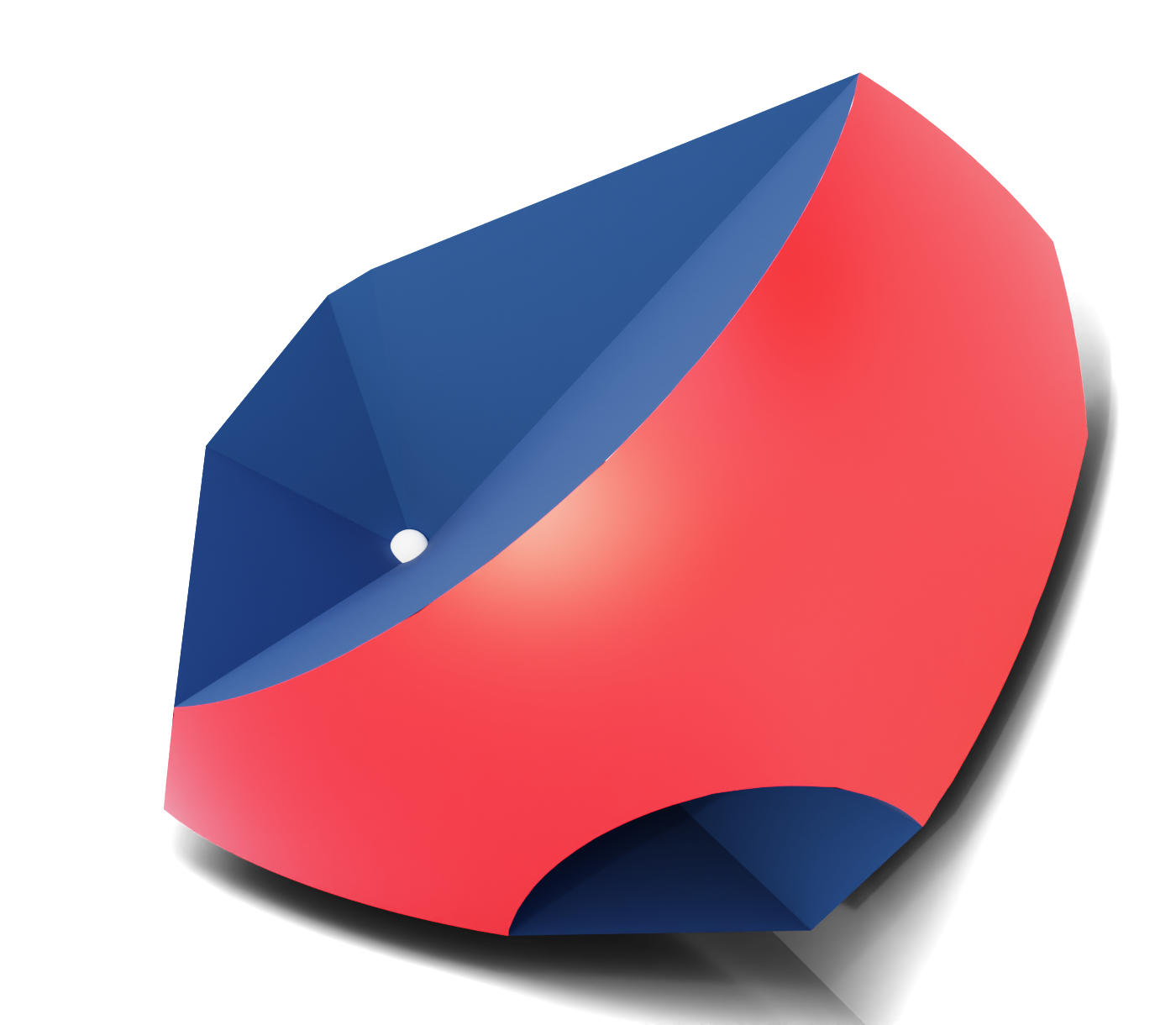}}\hfill
    \subfloat[Exploded-view]{\includegraphics[width=.47\columnwidth]{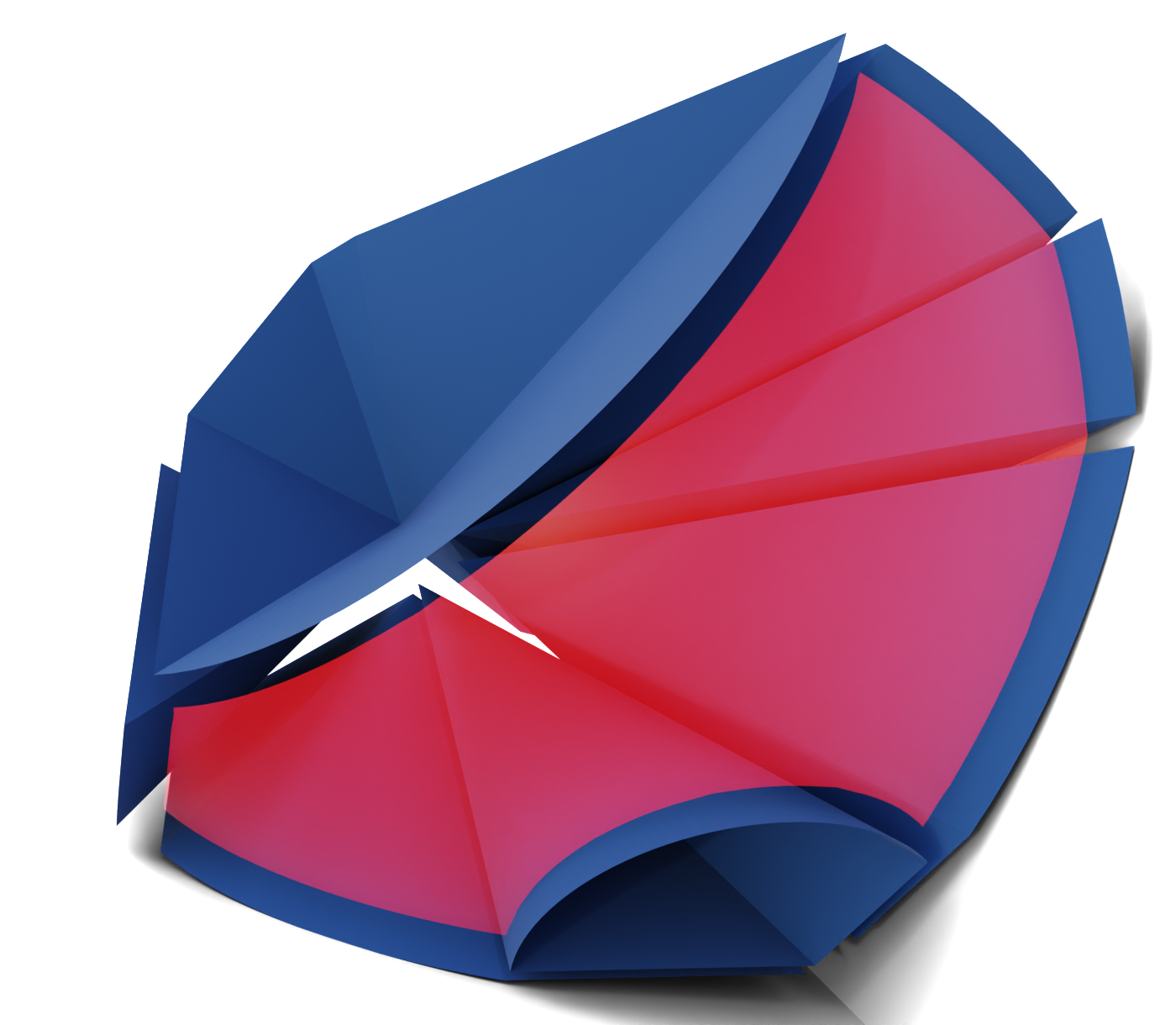}}
    \caption{Pyramid decomposition used for the computation of the differential quantities. Pyramids are defined by an apex located on the site (white) and a base which is a restricted Laguerre facet (blue) or a part of the spherical shell (red).}
    \label{fig:pyramid-decomposition}
\end{figure}

\subsection{Computation of Hessian and Gradient}
\label{ssec:hessian-gradient}

The computation of the unique weight vector satisfying the volume constraints is obtained as a solution of a convex optimization problem that can ben solved by a Newton algorithm (\Cref{ss:ot}). As shown by previous works~\cite{Levy2018, Kitagawa2019, Levy2022}, the gradient and Hessian of $K$ can be expressed in terms of the geometry of the Laguerre diagram. The gradient of the Kantorovich's functional is given by: \[
    \frac{\partial K}{\partial \psi_i} = \prescribedvolume_i - \abs{\powercell_i(\psi_i)}
\] while the coefficients of its Hessian are given by: \begin{equation}\begin{aligned}
    \frac{\partial^2 K}{\partial \psi_i \partial \psi_j} &=\frac{1}{2}\frac{|\powerbisector_{ij}(\weightset)|}{\norm{\site_j - \site_i}} && \text{if } i \neq j\\
    \frac{\partial^2 K}{\partial^2 \psi_i} &= -\left(\sum_{\site_j \in \mathcal{N}_i}{\frac{\partial^2 K}{\partial \psi_i \partial \psi_j}}\right) - \frac{1}{2}\frac{|\powerbisector_{i0}(\weightset)|}{\sqrt{\weight_i}} && \text{otherwise,}\label{eq:hessian}
\end{aligned}\end{equation} where $\powerbisector_{ij}$ denotes the Laguerre facet shared by the cells $\powercell_i(\weightset)$ and $\powercell_j(\weightset)$, $\mathcal{N}_i$ is the set of neighbors of site $\site_i$ that is, the set of sites $\site_j$ such that the cells $\powercell_i$ and $\powercell_j$ have a common facet $\powerbisector_{ij}$, and $\powercell_{i0}$ denotes the free surface associated with the cell $\powercell_{i}$. 

In a nutshell, the optimization process solving the optimal transport problem requires to compute the areas of the facets of the restricted Laguerre cells (including the curved ones adjacent to the background object $\object_0$) and their volumes. In the following sections, we show how these computations can be achieved analytically.

\subsection{Pyramid Decomposition}
\label{ss:pyramid-decomposition}

To compute the area of Laguerre facets and the volume of the restricted cells, one needs to know how they can be decomposed into simpler elements. While facets are planar polytopes which area can be simply obtained, restricted cells require to take into account complex boundaries with piecewise spherical shells.

\setlength{\intextsep}{0pt}%
\setlength{\columnsep}{0pt}%

\begin{wrapfigure}[6]{r}{.35\columnwidth}\centering\includegraphics[width=.3\columnwidth]{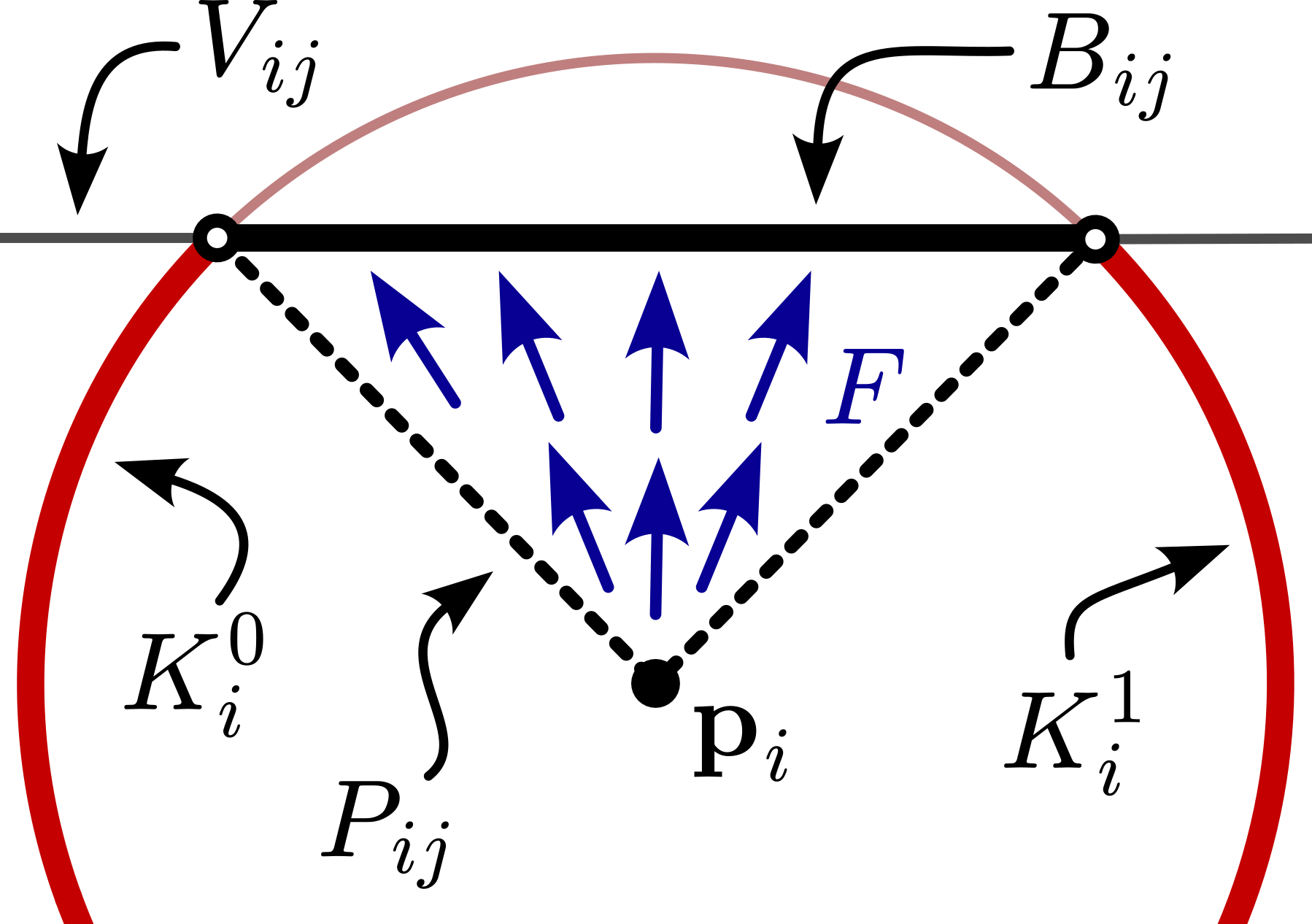}\end{wrapfigure}  This kind of problem has been met in various contexts and notably when studying chemical structures~\cite{Cazals2011, Duan2020}. In their work, \citeauthor{Cazals2011} have proposed decomposing them into a set of pyramids $P_{ij}$ which apex is the center of the sphere $\site_i$ and base $B_{ij}$ is a restricted facet $\powercell_{ij}$ or a spherical shell $K_{i}^t$ bound by several spherical circle, as illustrated in \Cref{fig:pyramid-decomposition}.

This decomposition has the advantage that the pyramid volume can be computed simply. Let $\mathbf{F} = \tfrac{(\point-\site_i)}{3}$ denote a vector field. It is easy to check that we have $\nabla \cdot \mathbf{F} = 1$ and then: \begin{align*}
    |P_{ij}| = \int_{P_{ij}}d\point = \int_{P_{ij}}\nabla \cdot \mathbf{F} \;d\point = \int_{\partial P_{ij}} \mathbf{F} \cdot n \;d\point,
\end{align*} with the last step obtained by applying the divergence theorem. We can notice that, by definition of $\mathbf{F}$, $\mathbf{F} \cdot n = 0$ on the sides of the pyramid, which gives: \[
    |P_{ij}| = h_{ij} \, |B_{ij}|,
\] where $h_{ij}$ the signed height of the pyramid given by: \[
h_{ij} = \begin{cases}
    \sqrt{\psi_i} & \text{if } B_{ij} \text{ is a spherical shell} \\
    \frac{\left(\norm{\site_j - \site_i}^2 + \psi_i - \psi_j\right)}{\norm{\site_j - \site_i}}(\site_j - \site_i) & \text{otherwise.}
\end{cases}
\]

Restricted Laguerre facets areas can then be computed by decomposing the generalized polygons into triangles and circular sectors. Then, the corresponding pyramid volume can be derived. However, handling spherical shells requires a robust method for computing their areas.

\subsection{Area of Spherical Patches}
\label{ss:spherical-patches}

\begin{wrapfigure}[6]{r}{.3\columnwidth}\centering\includegraphics[width=.25\columnwidth]{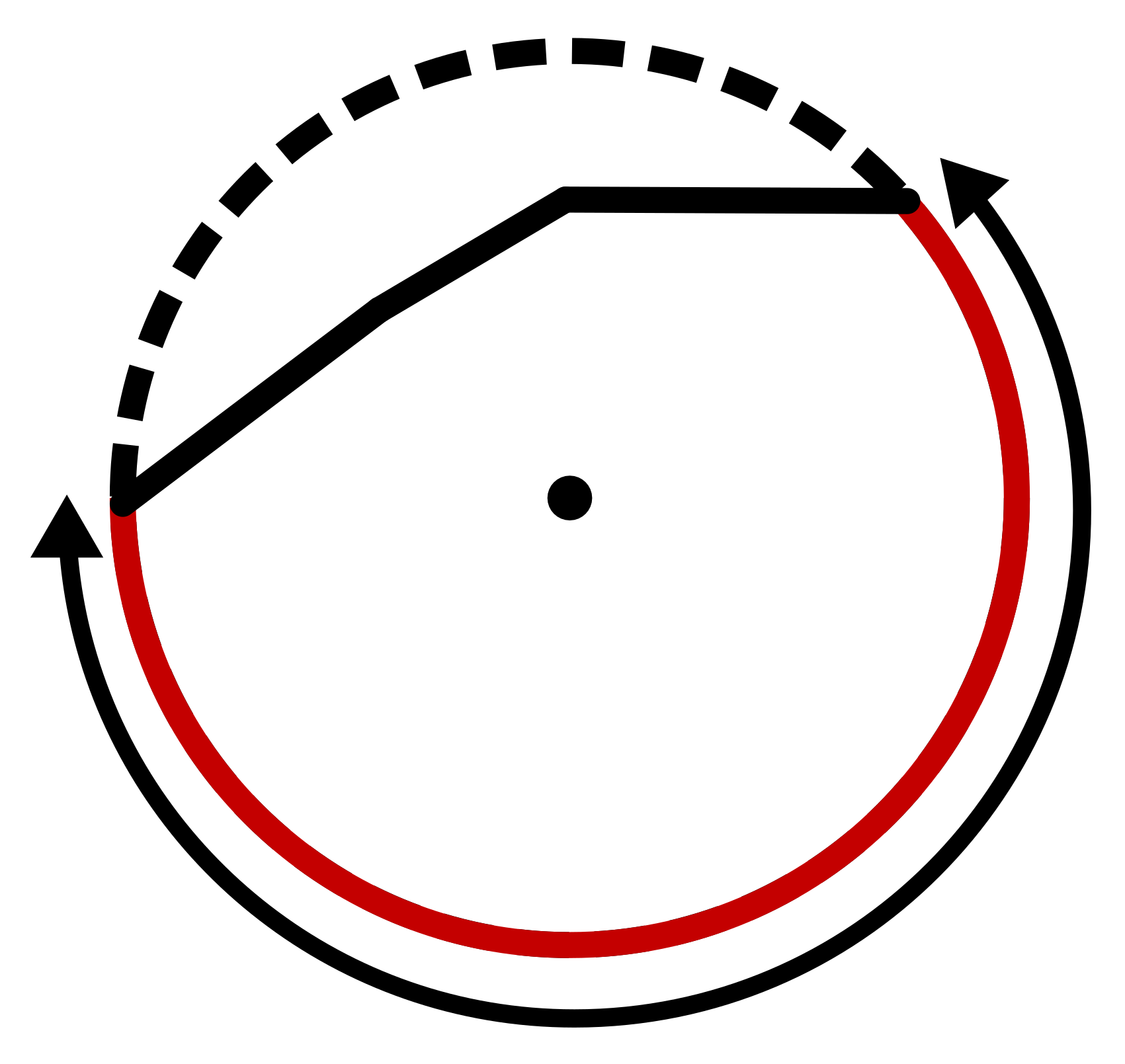}\end{wrapfigure}The restricted cell can be potentially composed of multiple spherical patches $K_{i}^t$. These patches are delimited by a set of spherical arcs or circles $b_{ij}$, derived from the restriction of Laguerre facets to the sphere, and vertices $v_{ijk}$, located at the intersection between a Laguerre edge and the sphere. Previous works \cite{Cazals2011, Duan2020}, rely on the Gauss-Bonnet theorem generalization to piecewise continuous manifolds to compute the area of these shells which is given by:\[
    \sum_{b_{ij} \in K_{i}^t}{k_g(b_{ij})} + \frac{|K_{i}^t|}{\psi_i} + \sum_{v_{ijk} \in K_{i}^t}\theta(v_{ijk}) = 2\pi \chi(K_{i}^t),
\] where $\theta(v_{ijk})$ is the angle at vertex $v_{ijk}$ between its two corresponding arcs bounding $K_{i}^t$, $\chi(K_{i}^t)$ is the Euler characteristic of the manifold, and finally, $k_g(b_{ij})$ is the geodesic curvature which is constant on a given circle:\[
    k_g(b_{ij}) = \frac{\sqrt{r_{ij}^2 - \weight_i}}{r_{ij} \sqrt{\weight_i}},
\] where $r_{ij}$ denotes the radius of the circle and $\weight_i$ the squared radius of the restriction sphere $\Sigma_i$.

\begin{wrapfigure}[6]{r}{.3\columnwidth}\centering\includegraphics[width=.25\columnwidth]{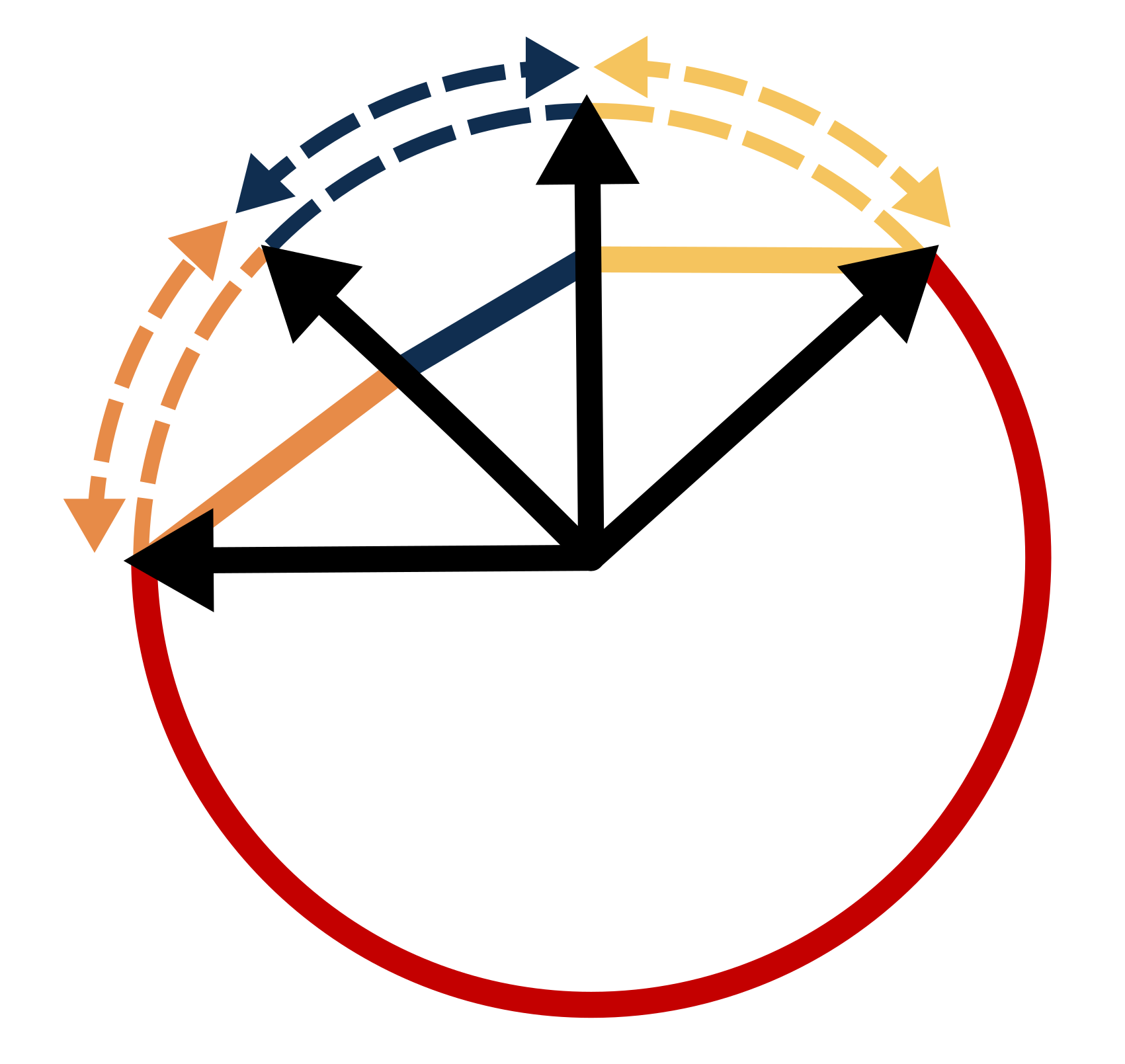}\end{wrapfigure}This formula requires to accurately compute the boundaries of each spherical patch. Previous works that construct these boundaries require cell-wise combinatorial information as well as exact predicates which are not always compatible with parallel processing. Besides their computational cost, they often require manipulating exact numbers\footnote{That do require dynamic memory allocation, which is incompatible with efficient parallel execution.}. To avoid these costly computation and data structure that would be difficult to implement on GPU, we propose instead an alternative method which only requires per-facet combinatorial information to efficiently compute the area of the spherical shells.

\begin{figure}[t]
    \centering
    \subfloat[Spherical patch and its boundary.]{\includegraphics[width=.49\columnwidth]{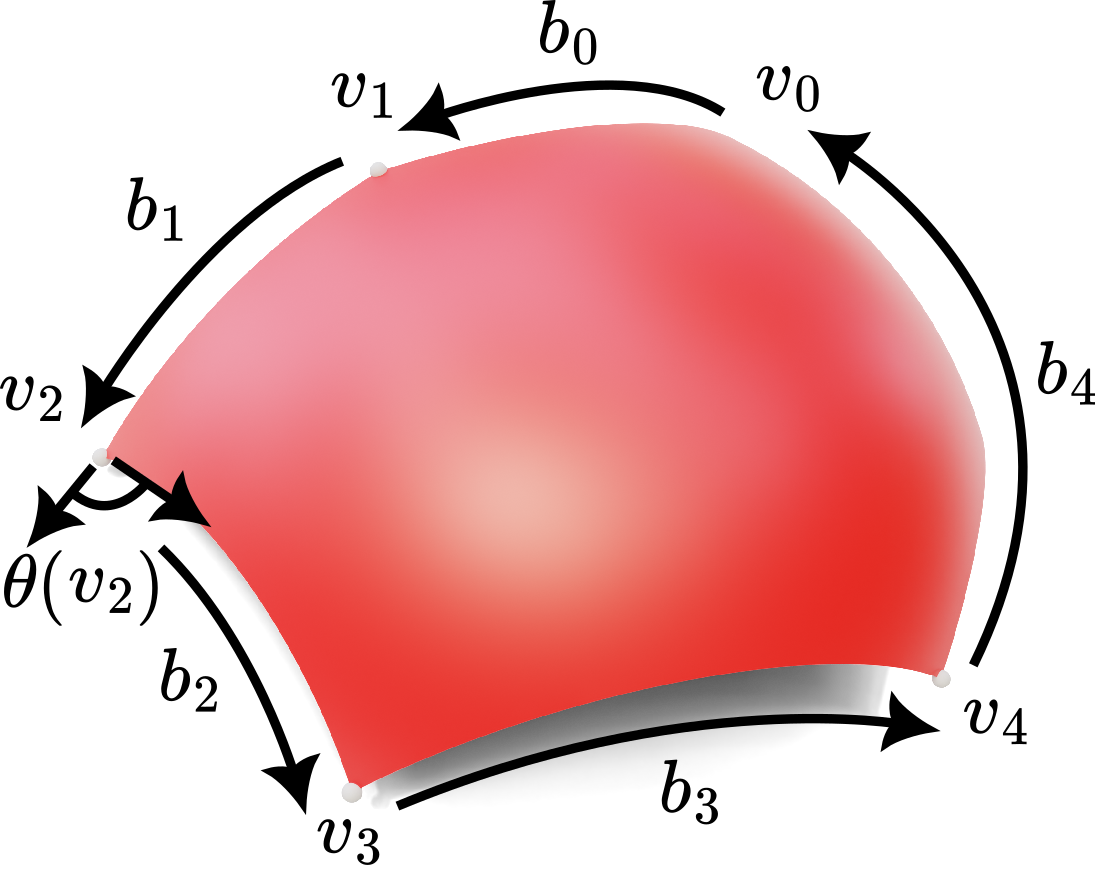}}\hfill
    \subfloat[Projection of a facet.]{\includegraphics[width=.49\columnwidth]{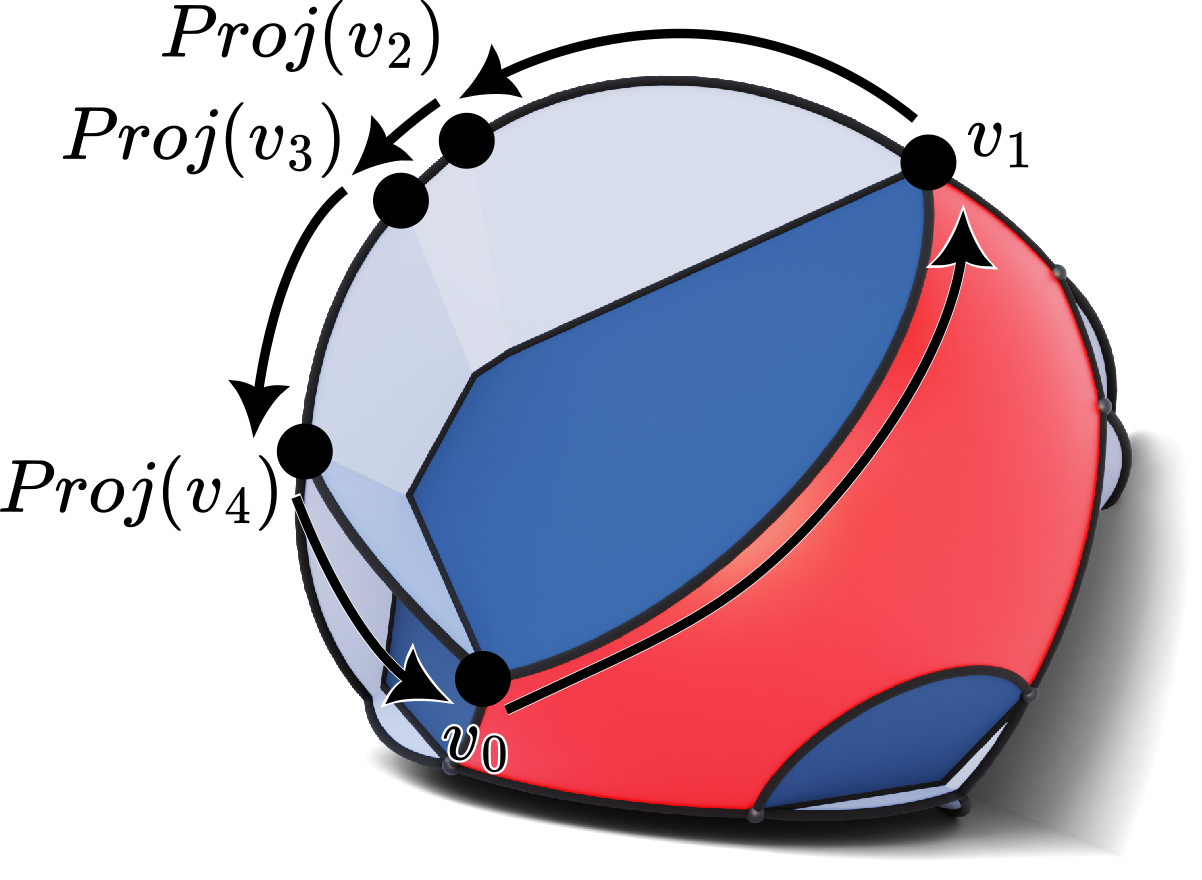}}
    \caption{Illustration of the area computation with spherical patches. (a) A spherical patch is bounded by arcs and vertices derived from the Laguerre cell and the sphere which may require exact predicates. (b) The area of projection of the cell facets onto the sphere can be computed without dedicated exact predicates.}
    \label{fig:area-computation}
\end{figure}

We can notice that, even if multiple spherical patches may be located on the same cell, only the \emph{sum of their area} $|K_i| = \sum_t |K_{i}^t|$ is actually needed for volume computation (\Cref{ss:pyramid-decomposition}). Additionally, this area can be expressed relative to the total area of the sphere $|\Sigma_i|$. Based on this observation, we propose to derive this quantity by subtracting the area covered by each restricted facet from the total area of the sphere (\Cref{fig:area-computation}). Let $c$ be a point located within the restricted cell, and $\text{Proj}(\powerbisector_{ij}, \Sigma_i, c)$, the projection operator of a bisector $\powerbisector_{ij}$ from $c$ onto the sphere $\Sigma_i$. Thus, we can define an \textit{occluded} sphere patch $\bar{K}_{ij}$ from the point of view of $c$: \[
    |\bar{K}_{ij}| = |\text{Proj}\left(\powerbisector_{ij}, \, \Sigma_i, \, c\right)|.
\] The total spherical patch area is then given by: \[
    |K_i| = |\Sigma_i| - \sum_{\powerbisector_{ij} \in \powercell_i}{|\bar{K}_{ij}|}.
\] This alternative formulation is solely based on the boundaries of each restricted bisector $\powerbisector_{ij}$ and does not need a sophisticated combinatorial representation.

Although our strategy does not require a dedicated exact predicate, we still need to find an adequate point $c$ located within the restricted cell for a non-degenerated numerical projection. Starting from each restricted facets centroid, we cast a ray which is intersected by the sphere and every other facet. We then average the center of each segment, resulting in a point inside the restricted cell.

\subsection{Restriction of a Laguerre cell to a sphere}
\label{ss:restriction}

Given a Laguerre cell which can be computed with previous works on a multi-core CPU~\cite{CGAL,geogram} or on a GPU~\cite{Basselin2021}, we need to find its restriction to the sphere $\Sigma_i$ to compute the quantities described in previous sections. 

\begin{algorithm}[!t]
    \DontPrintSemicolon

	\SetArgSty{upshape}
    \KwData{A Laguerre cell $\powercell_i$ of the site $\site_i$ and of weight $\weight_i$.}
    \KwResult{The set of restricted facet area $|B_{ij}|$, the free surface area $|K_i|$, and the restricted cell volume $|V_i|$.}

    \caption{Facets areas and volume of a Laguerre cell restricted to a sphere}\label{alg:restricted-quantities}

	$|V_i| \leftarrow 0$\;
    $\bar{|K_i|} \leftarrow 0$\;
    $c \leftarrow \text{point\_in($\powercell_i$)}$\;
    \ForAll{$\powerbisector_{ij} \in \powercell_i$}{
        $B_{ij} \leftarrow \text{restrict($\powerbisector_{ij},\, \Sigma_i$)}$ \Comment*{\textbf{\cref{ss:restriction}}}
        $\bar{|K_i|} \leftarrow \bar{|K_i|} + |\,\text{Proj}(B_{ij}, \;\Sigma_i,\; c)\,|$ \Comment*{\textbf{\Cref{ss:spherical-patches}}}
        
        $|P_{ij}| \leftarrow h_{ij} \; |B_{ij}|$ \Comment*{\textbf{\Cref{ss:pyramid-decomposition}}}
        $|V_{i}| \leftarrow |V_{i}| + |P_{ij}|$\;
    }
    
    $|K_i| \leftarrow |\Sigma_i| - \bar{|K_i|}$\Comment*{\textbf{\Cref{ss:spherical-patches}}}
    $|V_{i}| \leftarrow |V_{i}| + \sqrt{\weight_i}|K_i|$\Comment*{\textbf{\Cref{ss:pyramid-decomposition}}}
\end{algorithm}
 
\begin{wrapfigure}[5]{r}{.3\columnwidth}\centering\includegraphics[width=.25\columnwidth]{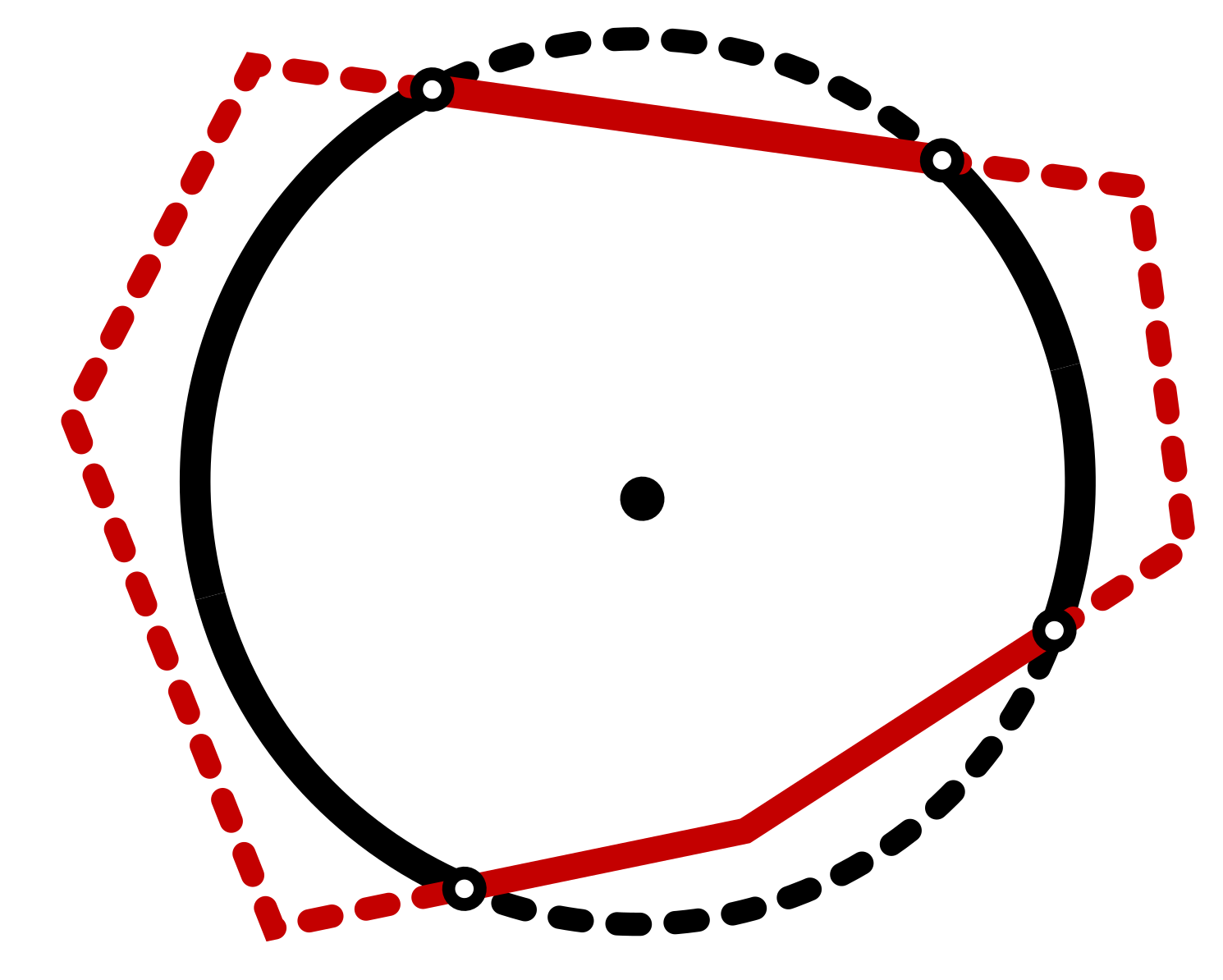}\end{wrapfigure}
Since the area of spherical patches is implicitly computed from our projection strategy, restricted facets represent the sole geometry needed for the optimization process. Based on this observation, we can reformulate this problem as the restriction of a simple polygon to its corresponding circle on the sphere. Laguerre facets are then individually processed and, based on the location of their vertices, restricted to the sphere. This is achieved by removing outer vertices and adding new vertices resulting from the intersection between an edge and the sphere. The final structure is composed of a set of facets of three types: polygonal facets, full circles, and generalized polygons including circular segments. Finally, we must also detect an empty intersection between the ball $\Sigma_i$ and the Laguerre cell $\powercell_i$ as well as the complete inclusion of $\Sigma_i$ into $\powercell_i$. If no facets are within $\Sigma_i$, the cell is a full sphere if the site $\site_i$ is within the cell and empty otherwise.

\begin{figure*}
    \subfloat[Prescribed cell volume of $2\mathrm{e}{-6}$ and initial configuration with $10 000$ sites.\label{fig:render-shower-10k}]{\includegraphics[width=.33\textwidth]{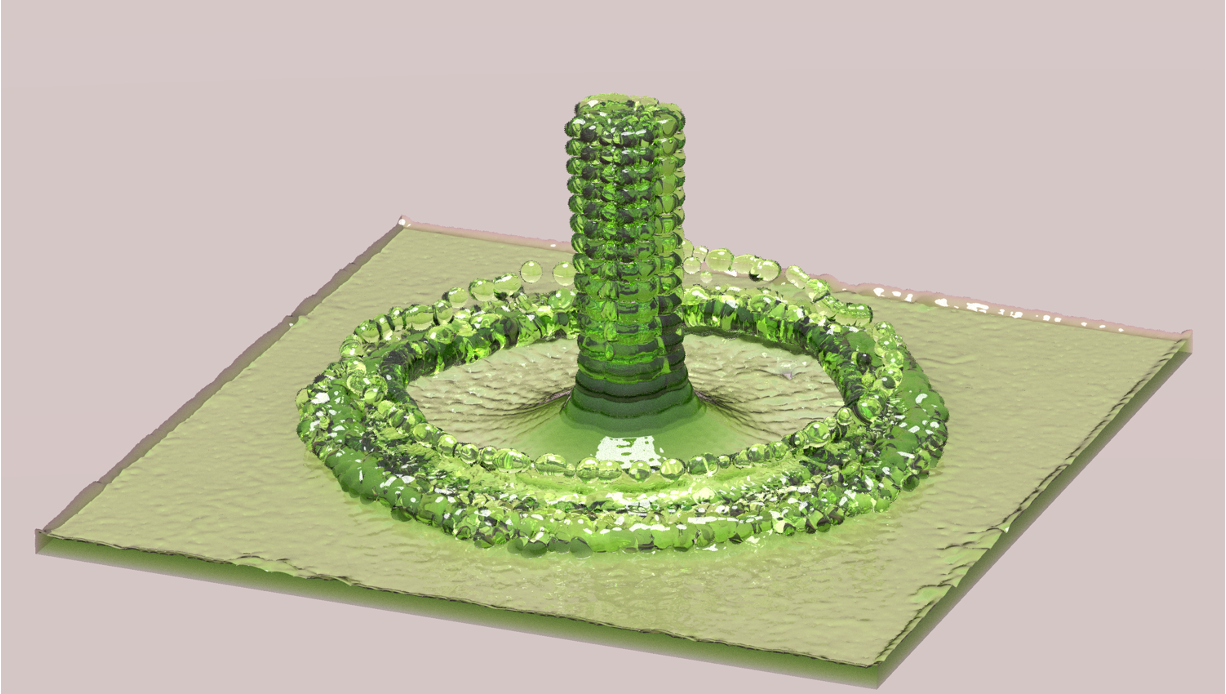} \hfill \includegraphics[width=.33\textwidth]{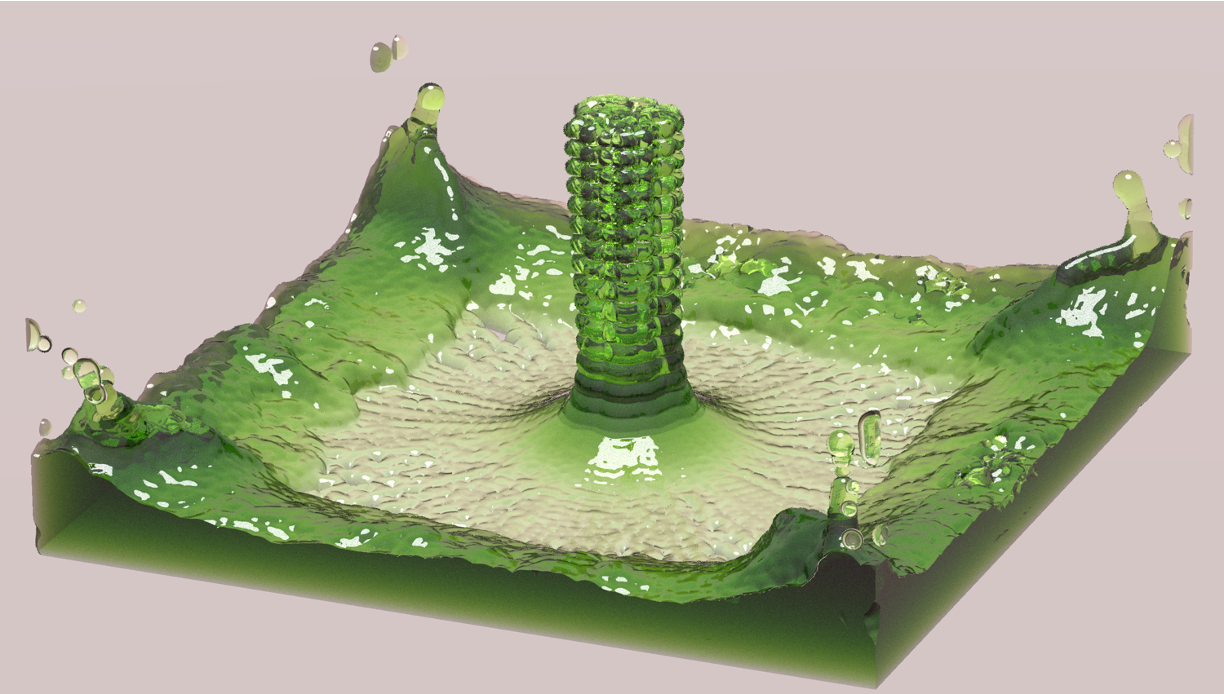} \hfill \includegraphics[width=.33\textwidth]{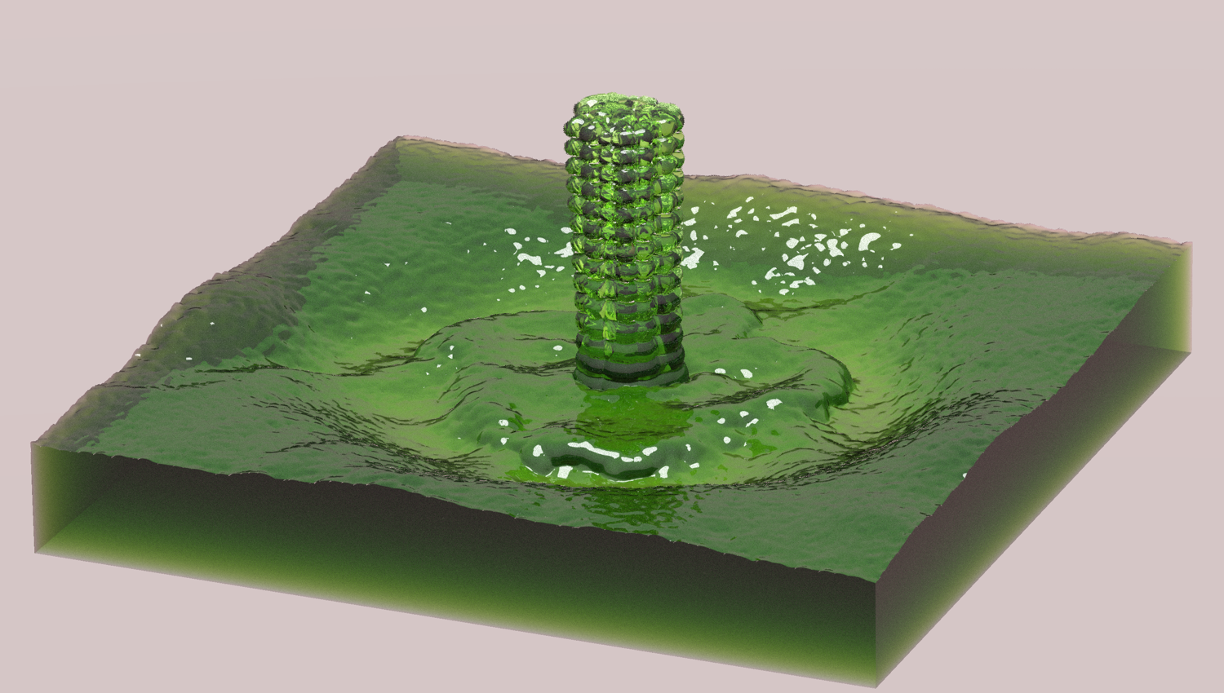}}\\
    \subfloat[Prescribed cell volume of $3\mathrm{e}{-7}$ and initial configuration with $50 000$ sites.\label{fig:render-shower-50k}]{\includegraphics[width=.33\textwidth]{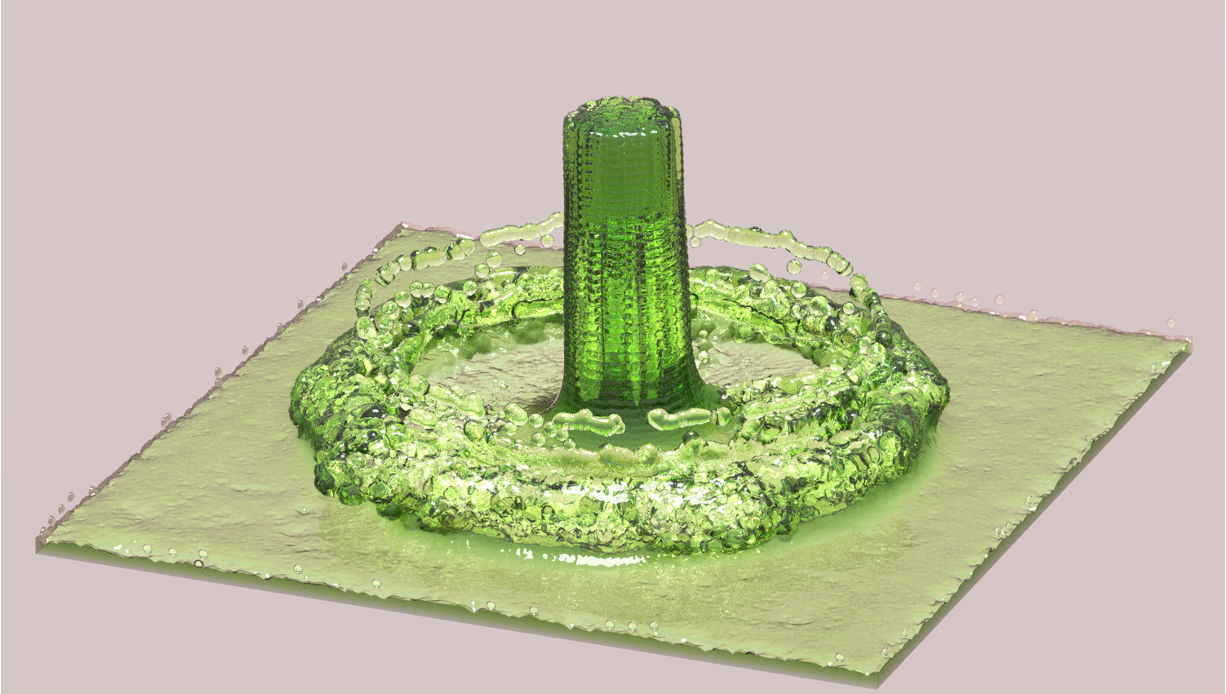} \hfill \includegraphics[width=.33\textwidth]{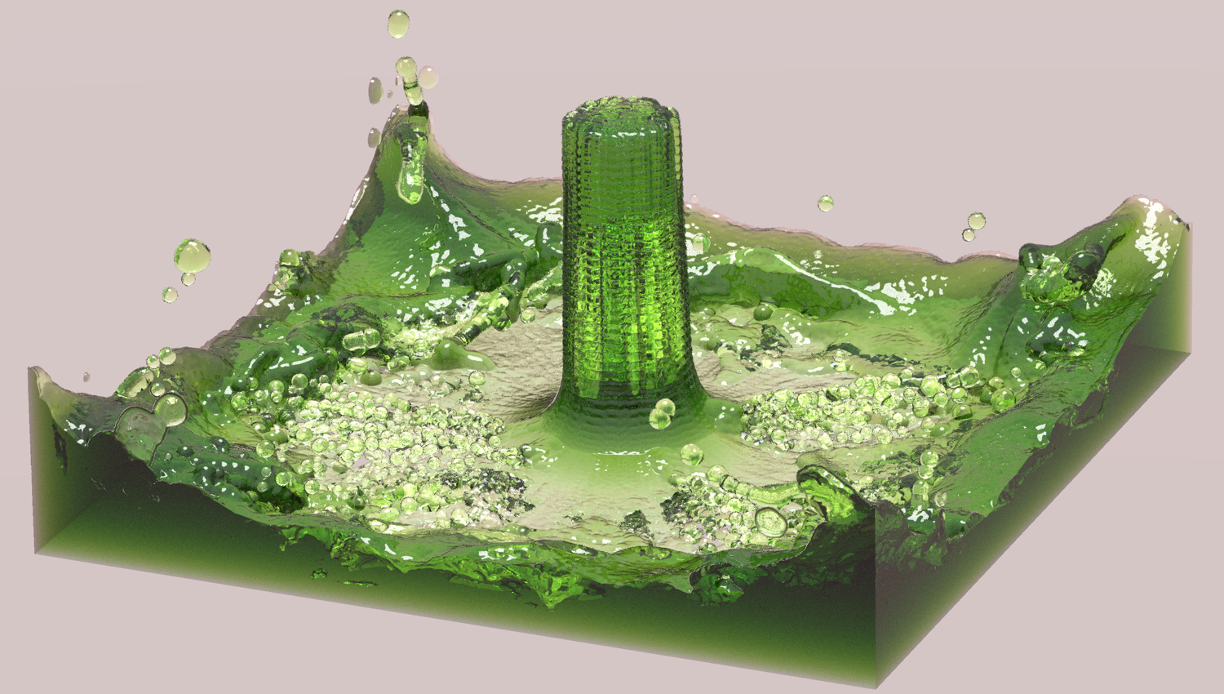} \hfill \includegraphics[width=.33\textwidth]{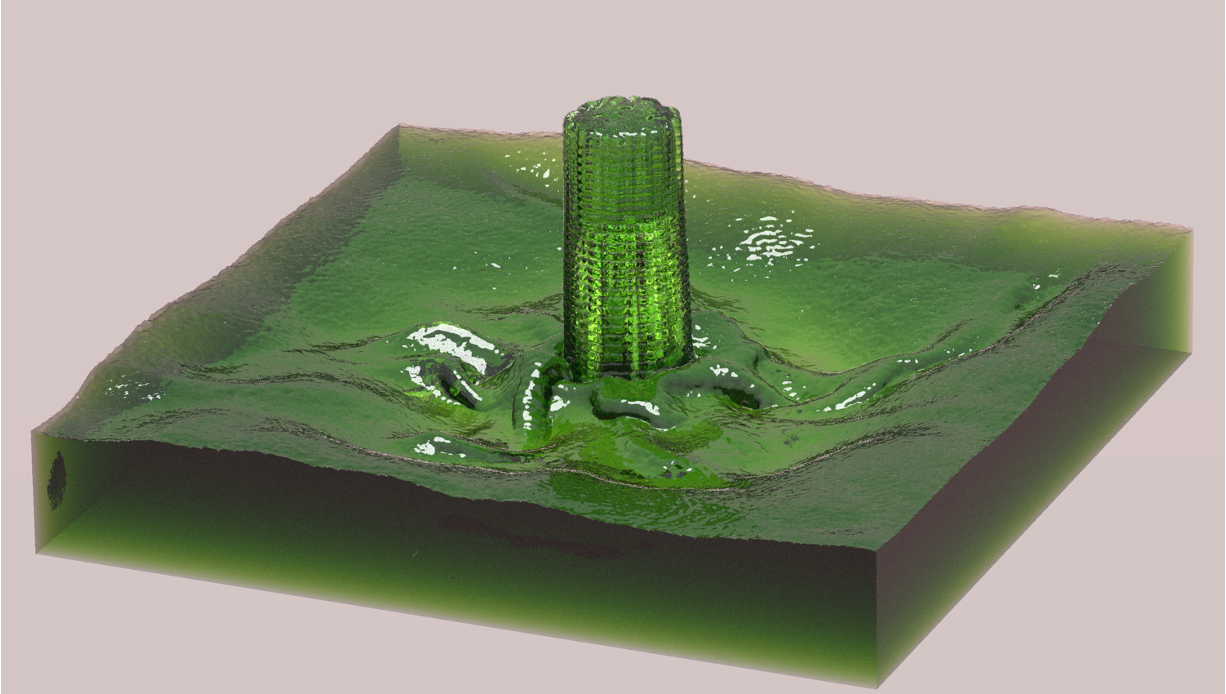}}
    \caption{A column launches a fast stream of fluid downwards. This produces detailed effects like droplets and complex waves while accurately preserving the volume. Our implementation robustly  handles the complex geometrical events produced during this simulation (top) even when the resolution is increased (bottom).}
    \label{fig:render-shower}
\end{figure*}

Our method to compute restricted cells, volumes and facets is summarized in Algorithm~\ref{alg:restricted-quantities}. By plugging these expressions into Algorithm~\ref{alg:volume-optimization} that computes the gradient and Hessian (\Cref{ssec:hessian-gradient}) of the objective function, it enables us to efficiently solve partial optimal transport problems. In the following sections, we describe how it can be used to perform the complete optimization on the GPU and efficiently render the fluid.

\section{GPU Implementation}
\label{sec:gpu}

By eliminating the needs for the triangle discretization of the spherical shells, our method enables a GPU implementation of the optimal transport solver. At its core, it requires two key components: the \emph{construction of Laguerre diagrams} and the \emph{evaluation of its differential quantities}.

Multiple previous works have been proposed to leverage the massively parallel architecture of GPU to build Laguerre diagrams from the geometry of their cell allowing independent construction~\cite{Ray2018, Basselin2021}. Each cell is initialized with the domain bounding box and iteratively clipped using bisectors defined by their set of nearest neighbors. However, this process is often limited to homogeneous spatial distribution that cannot always be guaranteed across the simulation. Notably, irregular cell shape across the diagram involves strongly varying update complexity and contributing neighbors that are far from the site. Since this can produce very long computation time or erroneous computation, we propose an alternative processing addressing these limitations.

\begin{wrapfigure}{r}{0.3\columnwidth}
    \begin{center}
        \includegraphics[width=.25\columnwidth]{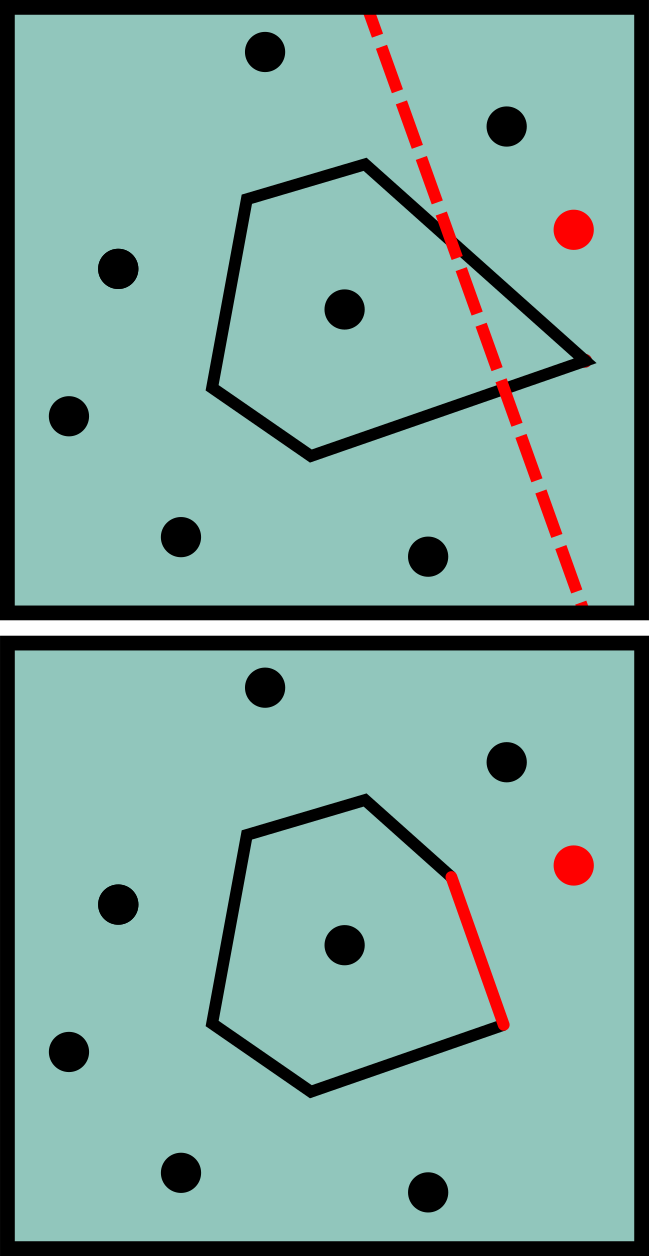}
    \end{center}
\end{wrapfigure}
\paragraph{Cell Update} Given the partially built cell of the site $\site_i$, the insertion of a (yet unconsidered) site $\site_j$ is made by clipping the geometry of the cell using the bisector $V_{ij}$. This is achieved by constructing the new facet based on the set of invalidated vertices. Previous works~\cite{Ray2018, Basselin2021} iterate over the cell structure through vertex adjacency which can only be made sequentially and requires robust combinatorics. Instead, we find clipped facets and sort their corresponding edges by their angle. New vertices can finally be directly computed from the sorted list of facets. 

\begin{wrapfigure}{r}{0.3\columnwidth}
    \begin{center}
        \includegraphics[width=.25\columnwidth]{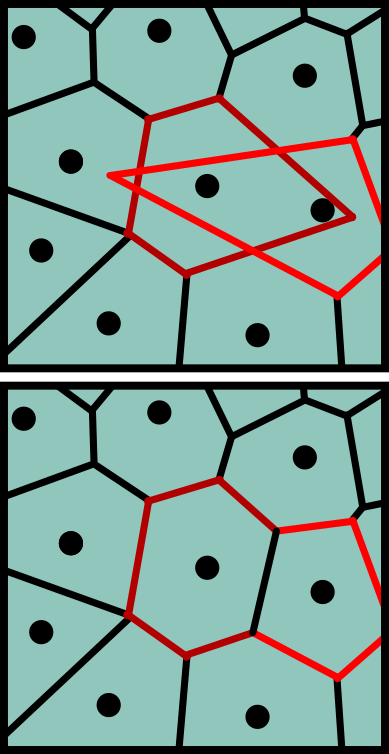}
    \end{center}
\end{wrapfigure}
\paragraph{Finding Contributing Neighbors} Previous works rely on the security radius~\cite{Ray2018, Basselin2021} to find the set of contributing neighbors by increasing distance order. However, the performance of this method is highly dependent on the assumption that the contributing set of sites is very close to the set of nearest neighbors, which cannot be satisfied during the optimization. In contrast, it is possible to find the contributing neighbors by noticing that, if two sites are contributing to each other's cell while not have been considered yet, the geometry of their cells intersects~\cite{Levy2025}. The construction then starts by initializing each cell with a few nearest neighbors allowing to build an acceleration structure from their geometry. All intersections are finally resolved by inserting all intersecting neighbors.

\paragraph{Restriction} Once Laguerre cells computed, one still needs to restrict cells and compute differential quantities. Thanks to our representation, this can be implemented easily on GPU by processing each face of the cell independently. Geometric data is recomputed on the fly when needed to reduce at most the memory cost in favor of computation, in line with GPUs strengths. 

\paragraph{Parallelism} Previous works~\cite{Ray2018, Basselin2021} process each cell with a single thread. Since cells are stored in shared memory, memory needs strongly increase with the number of threads in a single group. Additionally, when cells vary significantly, this produces highly heterogeneous computation needs and thus, divergence. To address these issues, we instead process each cell with a full warp. Memory can then be shared easily while the work can be dispatched more precisely. Our implementation not only takes advantage of warp-centric algorithms (\textit{e.g.} sorting algorithms for cell clipping) but also of warp intrinsic (\textit{e.g.} vote or register sharing).

\section{Fluid Mechanics}
\label{sec:physics}

Equipped with our optimal transport solver, we can now use it to implement our fluid simulation. Our representation inherits the properties of previous works~\cite{Goes2015,Levy2022}: \begin{itemize}
    \item through its Lagrangian point of view, it \emph{precisely tracks the velocity field} continuously across the simulation domain with \emph{strong guarantees regarding incompressibility};
    \item simulating \emph{volumes of fluids}, instead of particles, makes it possible to accurately compute differential quantities.
\end{itemize} In contrast with hybrid methods, maintaining two discretizations, this representation is unified. As observed in \cite{Goes2015}, it is simple to derive differential operators and use them to implement various physical effects. Building on previous works, we then describe our physical system including incompressibility, viscosity and surface tension. Applying forces on fluid volumes can only be performed indirectly through their sites, their only positional parameter. Special care must then be taken to accurately model fluid dynamics. 

\begin{figure*}
    \centering
    \includegraphics[width=\textwidth]{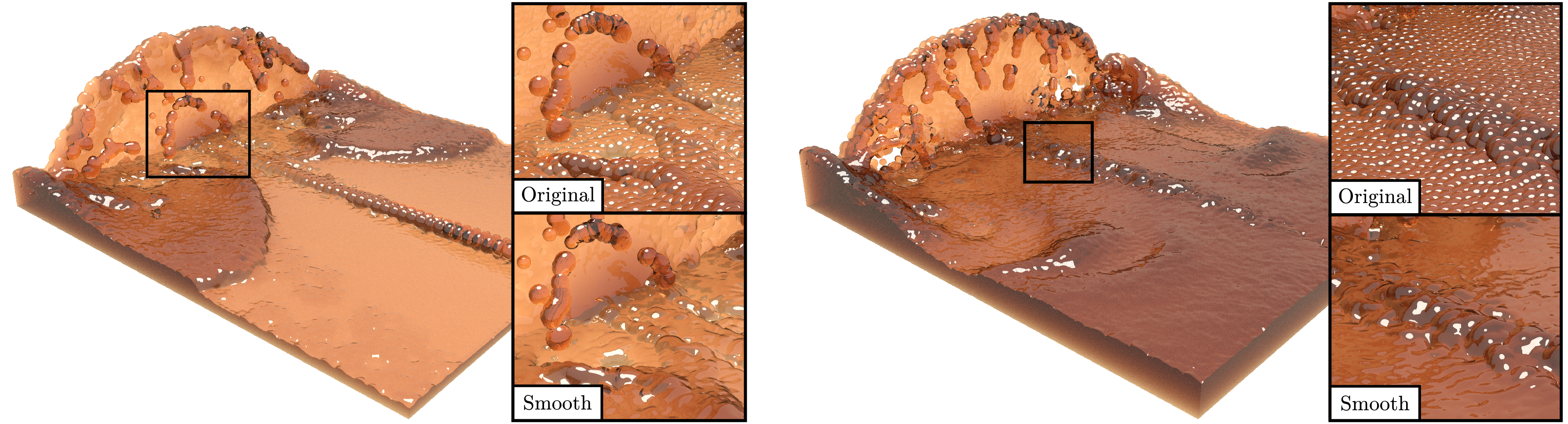}
    \caption{Two time steps of a fluid jet simulation initially composed of $10 000$ cells. A fluid with a high velocity collapses against the boundary and creates thin droplets. The closeup figures show the effect of surface smoothing as compared to raw renderings.}
    \label{fig:jet-10k}
\end{figure*}

\paragraph{Incompressibility} Even if we guarantee the preservation of cell volume, we must still compute an \emph{incompressible flow}. Initially, \citeauthor{Goes2015} projects the velocity to a divergent-free field by solving a Poisson equation~\cite{Goes2015}, as classically done in Lagrangian representations. Then, they move each site to its Laguerre cell barycenter. In contrast, \citeauthor{Gallouet2017} have proposed an alternative numerical scheme that provably converges to the Euler equations through the smooth projection of the velocity field onto the manifold of incompressible flow maps~\cite{Gallouet2017}. In a nutshell, pressure is modeled with a spring force between the site $\site_i$ and the centroid $\textbf{c}_i$ of its cell \[
    F_p = \tfrac{1}{\epsilon^2} (\textbf{c}_i - \site_i),
\] where $\epsilon$ is the strength of the spring. In the simulations presented throughout this article, we rely on \citeauthor{Gallouet2017}'s numerical scheme. Note that, as a ``plug and play'' solver for partial optimal transport problems, our method can also be applied to the \citeauthor{Goes2015} scheme that will benefit from the same improvement in both speed and accuracy.

\begin{wrapfigure}{r}{0.35\columnwidth}
    \begin{center}
        \includegraphics[width=.25\columnwidth]{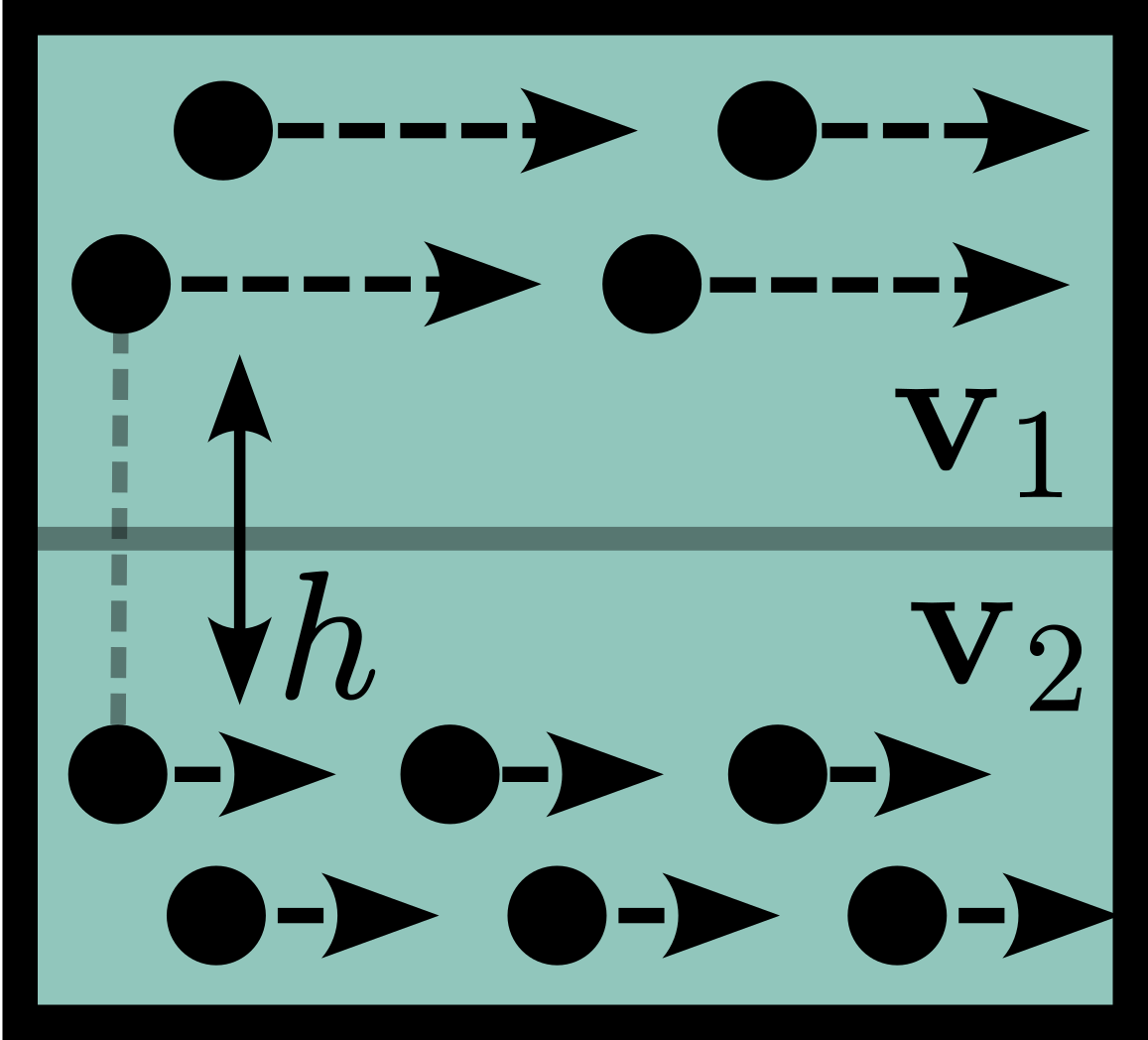}\\[1em]
        \includegraphics[width=.25\columnwidth]{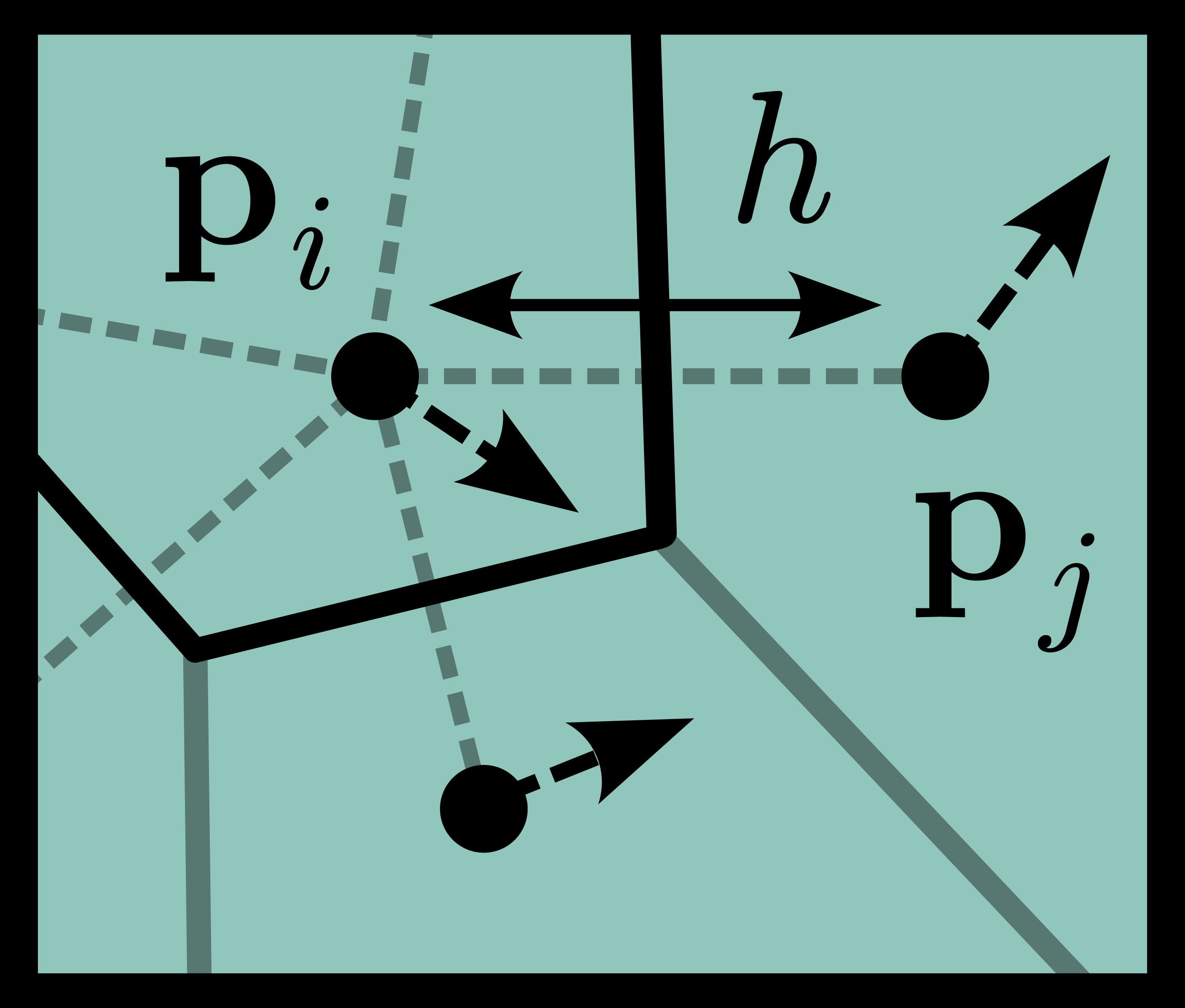}
    \end{center}
\end{wrapfigure}
\paragraph{Viscosity} Let us see now how to implement viscosity. One of the advantages of our Lagrangian implementation is that it can be intuitively derived, leading to a well-known formula (Laplacian of velocities). Consider two parallel streams of fluid with different velocities $\textbf{v}_1$ and $\textbf{v}_2$ (see inset) as well as two particles $\site_1$ and $\site_2$ in both streams separated by a distance $h$. Their frictional viscosity is a force that tends to homogenize their velocities, given by $F_v = \mu \tfrac{\textbf{v}_j - \textbf{v}_i}{h}$, where $\mu$ is a constant that depends on the physical properties of the fluids. Now consider a fluid volume $\prescribedvolume_i$, parameterized by point $\site_i$. It is influenced by the surrounding particles $\site_j$, each of them with an influence proportional to $|\powercell_{ij}|$, the area of the facet common to cells $\powercell_{i}$ and $\powercell_{j}$, and inversely proportional to the distance $\|\site_j - \site_i\|$ between both particles. Summing over all neighbors $\site_j$, one retrieves the classical formula of the $\mathbb P_1$ Finite Element Laplacian, also called cotan-weights Laplacian: \begin{equation} 
    F_v = \mu \hat{\Delta} \textbf{v} = \mu \left(\sum_{\site_j \in \mathcal{N}_i}{\frac{1}{2}\frac{|\powerbisector_{ij}(\weightset)|}{\norm{\site_j - \site_i}}} \right)\left(\textbf{v}_j - \textbf{v}_i\right).\label{eq:laplacian} 
\end{equation} Now consider the more general case of several fluids and fluid-boundary interaction (further explained later). Then, our empirical viscosity force $F_v$ is parameterized by $\mu_{ij}$ coefficients that depend on the natures of both fluids boundaries \[
    F_v = \left(\sum_{\site_j \in \mathcal{N}_i}{\frac{\mu_{ij}}{2}\frac{|\powerbisector_{ij}(\weightset)|}{\norm{\site_j - \site_i}}}\right) \left(\textbf{v}_j - \textbf{v}_i\right).
\]

\begin{figure}[!b]
    \begin{center}
        \subfloat[Concave meniscus $\mu_{ij} = 2.5$]{\includegraphics[width=.45\columnwidth]{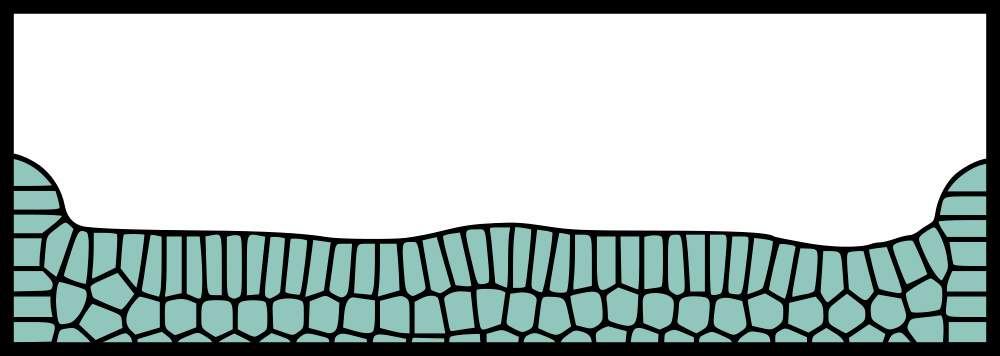}}
        \hspace{.5em}
        \subfloat[Convex meniscus $\mu_{ij} = 0.1$]{\includegraphics[width=.45\columnwidth]{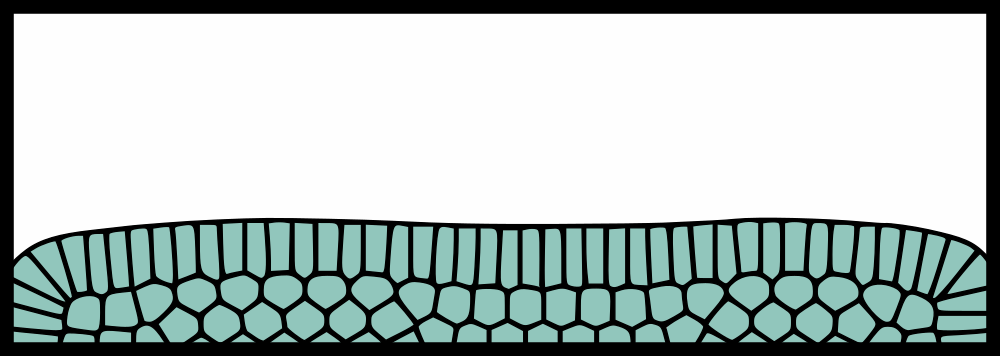}}
    \end{center}
    \caption{Meniscus obtained as an effect of surface tension that embeds fluid-boundary interaction.}
\end{figure}

\begin{wrapfigure}{r}{0.3\columnwidth}
    \begin{center}
        \setcounter{subfigure}{0}
        \subfloat[Bulk]{\includegraphics[width=.225\columnwidth]{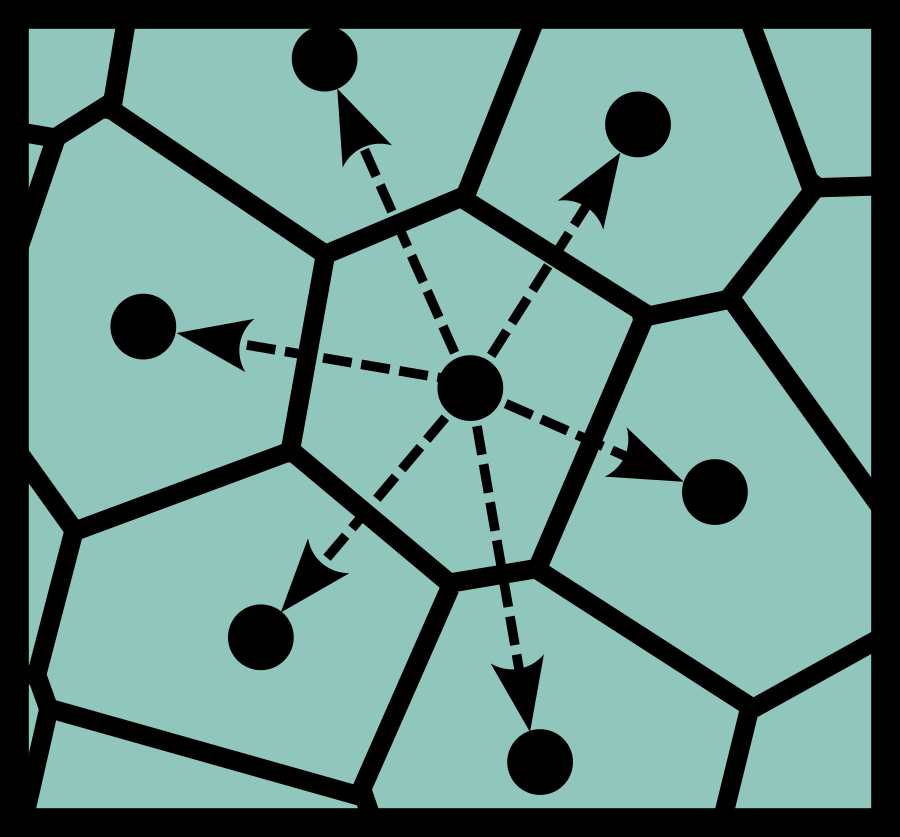}}\\
        \subfloat[Surface]{\includegraphics[width=.225\columnwidth]{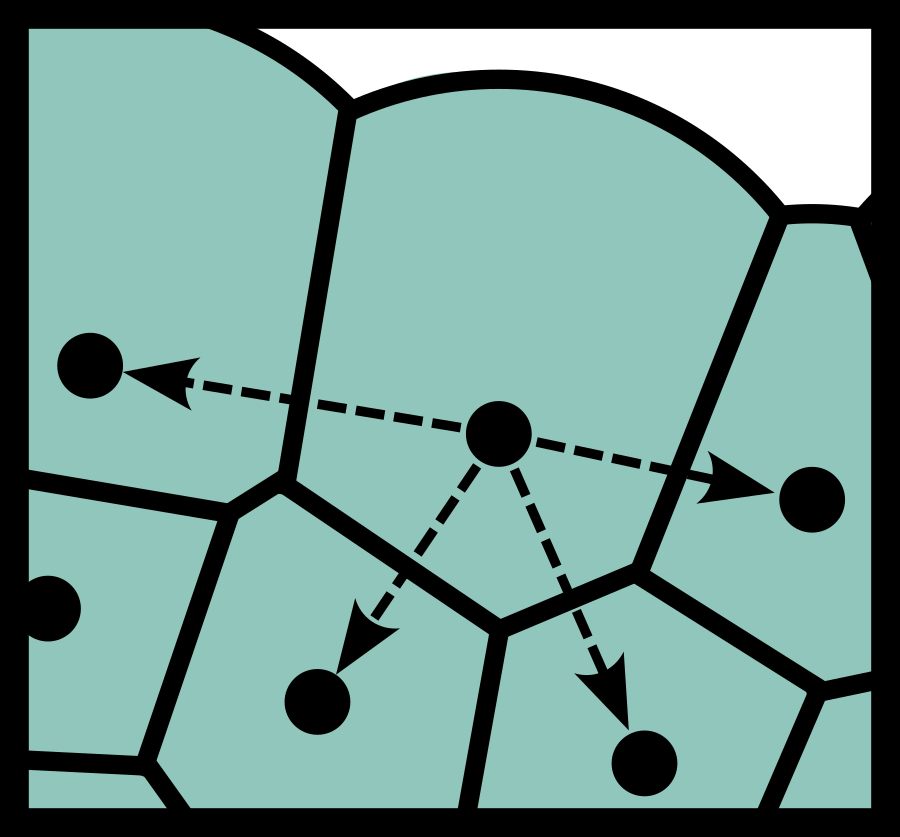}}
    \end{center}
\end{wrapfigure}
\paragraph{Surface tension} In the bulk of the fluid, the attractions of a particle by its surrounding neighbors cancel out. In contrast, near the air-fluid interface the asymmetry results in a net effect pulling the particle towards the interior of the fluid. More precisely, this tends to minimize the surface area of the fluid in a way that can be expressed as a relation between surface Gaussian curvature and the gradient of the pressure field, called the Young-Laplace equation~\cite{Das2019}. Since computing surface differential quantities like Gaussian curvature would require in principle a smooth ($C^2$) surface, we prefer instead to rely on a volumetric expression. Similarly to viscosity, surface tension can be expressed as an homogenization of the particle's locations within the volume $F_t = \gamma \Delta \site$ which can be easily discretized with, again, the $\mathbb P_1$ Laplacian $\hat{\Delta}$. Note that in this case, $\hat{\Delta}$ is the vector Laplacian which is the scalar Laplacian operator applied to each coordinate of $\site$. The coefficient $\gamma$ controls the strength of this surface tension force. Note that our expression of surface tension as a \emph{volumetric} effect avoids the need of evaluating curvatures on the interfaces.

\begin{figure*}
    \subfloat[Falling teapot containing a fluid droplet with varying viscosity levels indicated by color. The viscous cap falls onto the liquid droplet and creates a large fluid splash. (Yellow) Very low viscosity. (Pink) Medium viscosity. (Blue) Very high viscosity.\label{fig:teapot}]{\includegraphics[width=\textwidth]{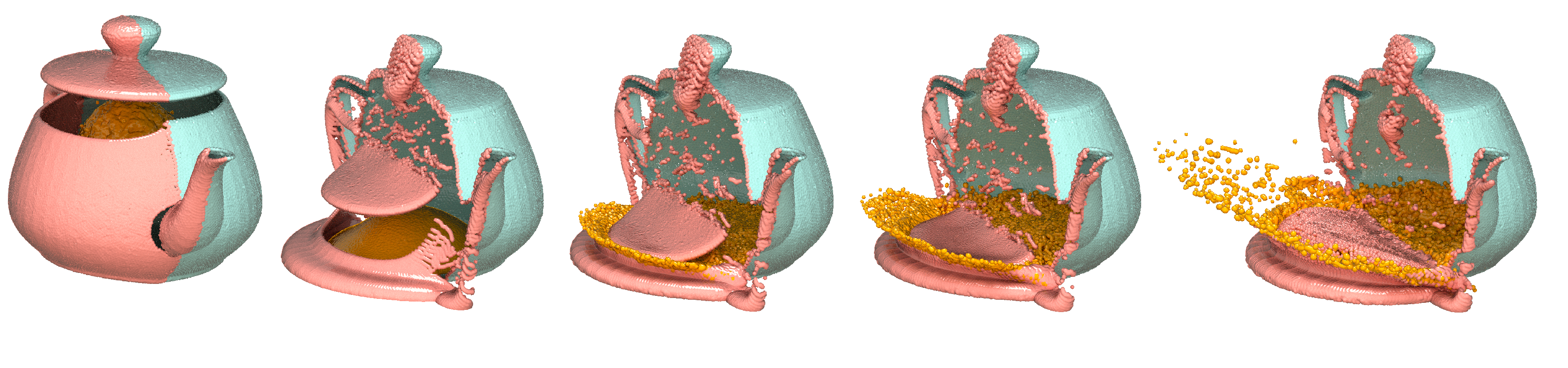}}\\
    \subfloat[Liquid bunny in zero gravity. Due to surface tension, the main force in such setting, fluid's shape slowly converges to a sphere.\label{fig:bunny-drop}]{\includegraphics[width=\textwidth]{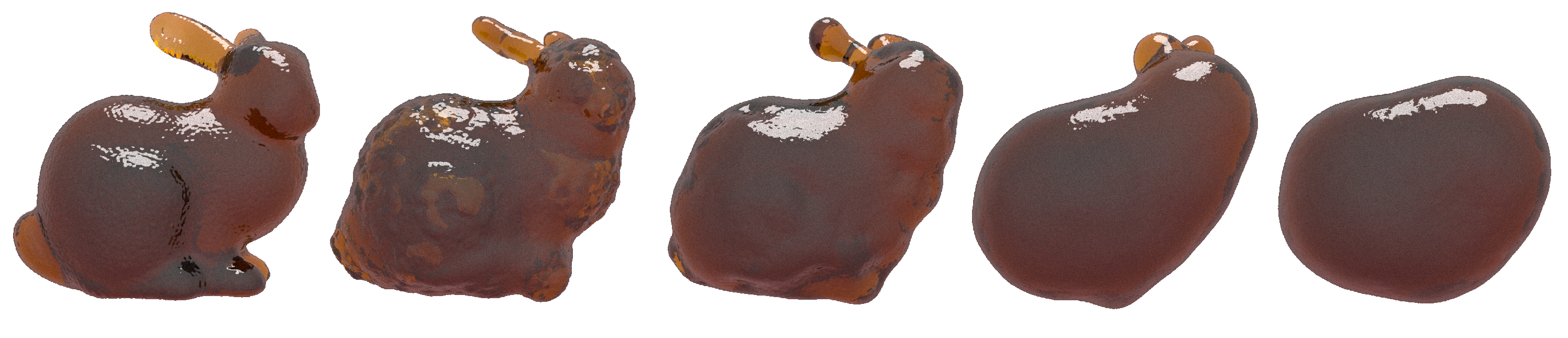}}
    \caption{Illustrations of two animations featuring complex physical effects handled by our framework. The interface between two fluid cells with different parameters can be handled accurately thanks to the unified representation of the geometry. Thanks to this property, we are able to accurately simulate fluids with varying configurations.}
    \label{fig:physical-effects}
\end{figure*}

\paragraph{Numerical integration} As previous works~\cite{Stam1999, Levy2022}, we rely on a semi-implicit integration ensuring unconditionally stable viscosity. Let $\textbf{x}^{k}$ and $\textbf{v}^{k}$ the position and velocity vectors at time step $k$, $\textbf{x}^{k+1}$ and $\textbf{v}^{k+1}$ are given by \[
    \begin{aligned}
        \textbf{x}^{k+1} &= \textbf{x}^{k} + \delta t \textbf{v}^{k}\\
        \textbf{v}^{k+1} &= \textbf{v}^{k} + \tfrac{\delta t}{m} (F_p + F_g + F_t + \mu\hat{\Delta}\textbf{v}^{k+1}),\\
    \end{aligned}
\] where $F_p$ denotes the incompressibility force, $F_g$ the gravity force and $F_t$ the surface tension force. Note that the viscosity $\mu \hat{\Delta}\textbf{v}^{k+1}$ in the right-hand side depends on the velocity at time step $k+1$. Hence $\textbf{v}^{k+1}$ is obtained as \emph{the solution of a linear system} \begin{equation}
    (\mu \hat{\Delta} - \tfrac{m}{\delta t} Id) \textbf{v}^{k+1} = -(\tfrac{m}{\delta t} \textbf{v}^{k} + F_p + F_g + F_t).\label{eq:semi-implicit-viscosity}
\end{equation}

\begin{wrapfigure}[10]{r}{0.4\columnwidth}
  \begin{center}
    \includegraphics[width=0.35\columnwidth]{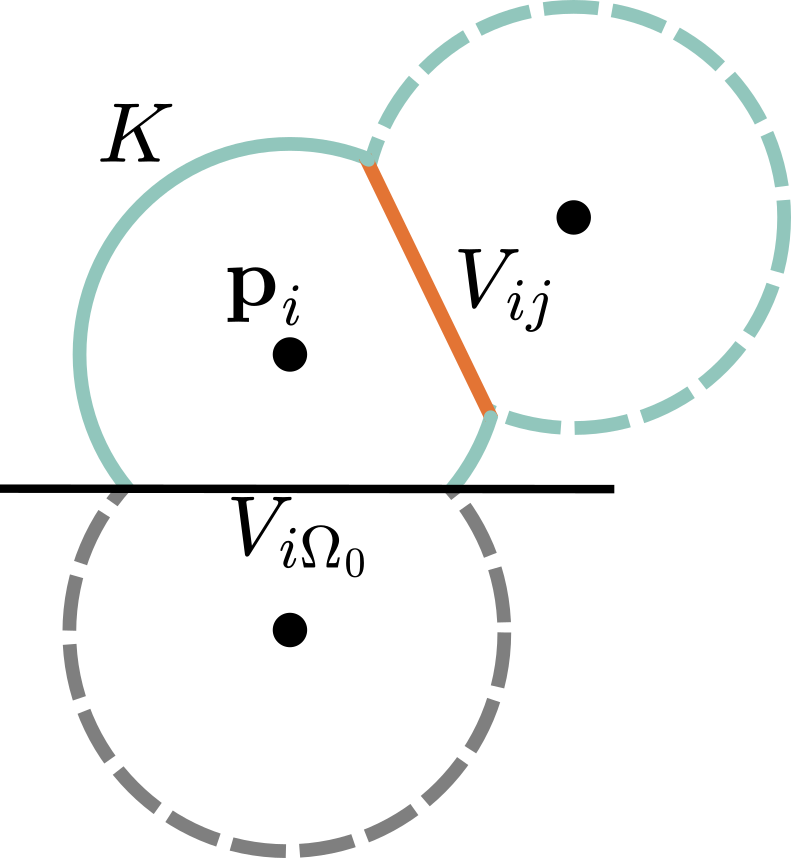}
  \end{center}
\end{wrapfigure}
\paragraph{Fluid-boundary interaction} Special care must be taken to accurately simulate the effect at fluid-boundary interfaces. For instance, a viscous fluid can \emph{stick} to a solid boundary while a fluid has a convex meniscus on a hydrophobic surface. Both effects are modeled by the $\mathbb P_1$ Laplacian, involved in the surface tension and viscosity, which must include boundary contributions of the free surface and the domain. Following \eqref{eq:laplacian}, the contribution of the domain boundary is simply given by\[
    \sum_j \frac{1}{2}\frac{\mu_{ij}|V_{i\Omega_j}|}{d(\site_i, \Omega_j) |V_i|},
\] where $\mu_{ij}$ denotes the affinity of fluid $i$ to boundary $j$. While we assume zero velocity on the boundary, we compute a corresponding position for surface tension at $\sqrt[3]{|V_i|}$ away from the bisector between the site $\site_i$ and the boundary $\Omega$. This position guarantees that each site with the same prescribed volume will have the same boundary weighting.

\section{Rendering}
\label{sec:rendering}

To efficiently display the fluid and compute complex lighting effects, we introduce a ray-tracing-based strategy relying on the presented volumetric description. For this purpose, two features are required: finding the starting cell of each pixel and performing the traversal of the fluid volume.

\begin{wrapfigure}{r}{0.275\columnwidth}
  \begin{center}
    \includegraphics[width=0.225\columnwidth]{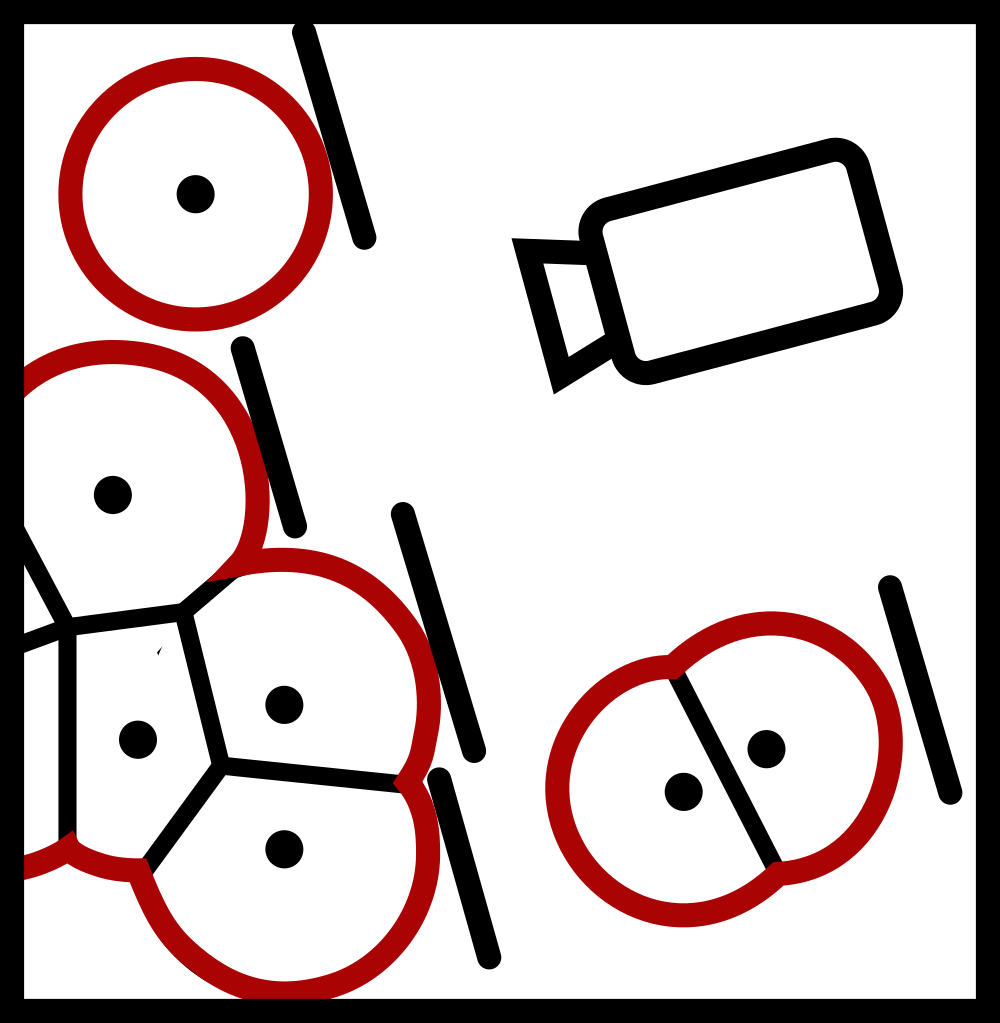}
  \end{center}
\end{wrapfigure}
\paragraph{Finding the First Cell} Starting the traversal of the geometric structure requires to find an entry point for the ray. This could be achieved with, \textit{e.g.} a nearest neighbor query relying on an additional acceleration structure. However, this would involve construction and traversal costs with regards to both memory consumption and performance. Thus, we prefer to only rely on the already computed restricted Laguerre Diagram and follow a similar strategy presented by previous works for \emph{SPH} rendering~\cite{Xiao2018}. To efficiently extract the free surface, authors project particles onto the screen by using impostors. In our context, only cells with spherical patches must be projected, reducing the cost. Once rasterized, each sphere is intersected with the ray in the fragment shader allowing a pixel-accurate rendering~\cite{Gralka2023}.

\paragraph{Volume and Empty Space Traversal} Once the first cell has been found, we can perform the actual traversal of the volume. Considering first a volume fully filled by a Laguerre diagram. To perform the traversal through cells, we can load the current cell's neighbors and iteratively intersect facets facing the ray, as noticed by previous works~\cite{Govindarajan2025}. The process can then be restarted with the neighbor of the closest facet until the closest facet is the domain boundary. Considering now a fluid with the free surface, we explore the empty part of the domain through the unrestricted Laguerre diagram. Depending on whether the traversal is currently within the fluid or the air, we limit the evaluation to facets between fluid cells, accelerating the traversal.

\begin{table*}
    \centering
    \begin{tabular}{l@{\hspace{20pt}} rrr|rrr|rrr|rrr}
    \toprule
    \multirow{2}{*}{\textbf{Simulation}} & \multicolumn{3}{c}{\textbf{Laguerre}} & \multicolumn{3}{c}{\textbf{Evaluation}} & \multicolumn{3}{c}{\textbf{Solve}} & \multicolumn{3}{c}{\textbf{Complete Step}} \\\cmidrule(lr){2-4}\cmidrule(lr){5-7}\cmidrule(lr){8-10}\cmidrule(lr){11-13}
     & CPU & GPU & Ratio & CPU & GPU & Ratio & CPU & GPU & Ratio & CPU & GPU & Ratio \\
    \midrule
    \Cref{fig:render-shower-10k} & 0.662 & 0.335 & 1.979 & 1.043 & 0.288 & 3.625 & 0.162 & 0.049 & 3.326 & 1.921 & 0.750 & 2.561 \\
    \Cref{fig:render-shower-50k} & 5.603 & 4.779 & 1.172 & 11.051 & 3.887 & 2.843 & 2.748 & 0.230 & 11.975 & 19.850 & 9.774 & 2.031 \\
    \Cref{fig:jet-10k} & 0.673 & 0.315 & 2.134 & 1.063 & 0.368 & 2.888 & 0.138 & 0.052 & 2.655 & 1.935 & 0.823 & 2.350 \\
    \Cref{fig:teapot} & 1.855 & 1.239 & 1.497 & 5.268 & 1.409 & 3.740 & 2.744 & 0.262 & 10.457 & 9.963 & 3.346 & 2.978 \\
    \Cref{fig:bunny-drop} & 0.957 & 0.916 & 1.045 & 2.395 & 0.708 & 3.381 & 0.506 & 0.044 & 11.369 & 3.880 & 1.844 & 2.104 \\
    \bottomrule
    \end{tabular}\\[1em]
    \caption{Per step mean timings of the presented animations relying on our two implementations. Values are given in seconds and are computed as the mean of the first $100$ simulation steps.}
    \label{tab:performance}
\end{table*}

\begin{wrapfigure}{r}{0.275\columnwidth}
  \begin{center}
    \includegraphics[width=0.225\columnwidth]{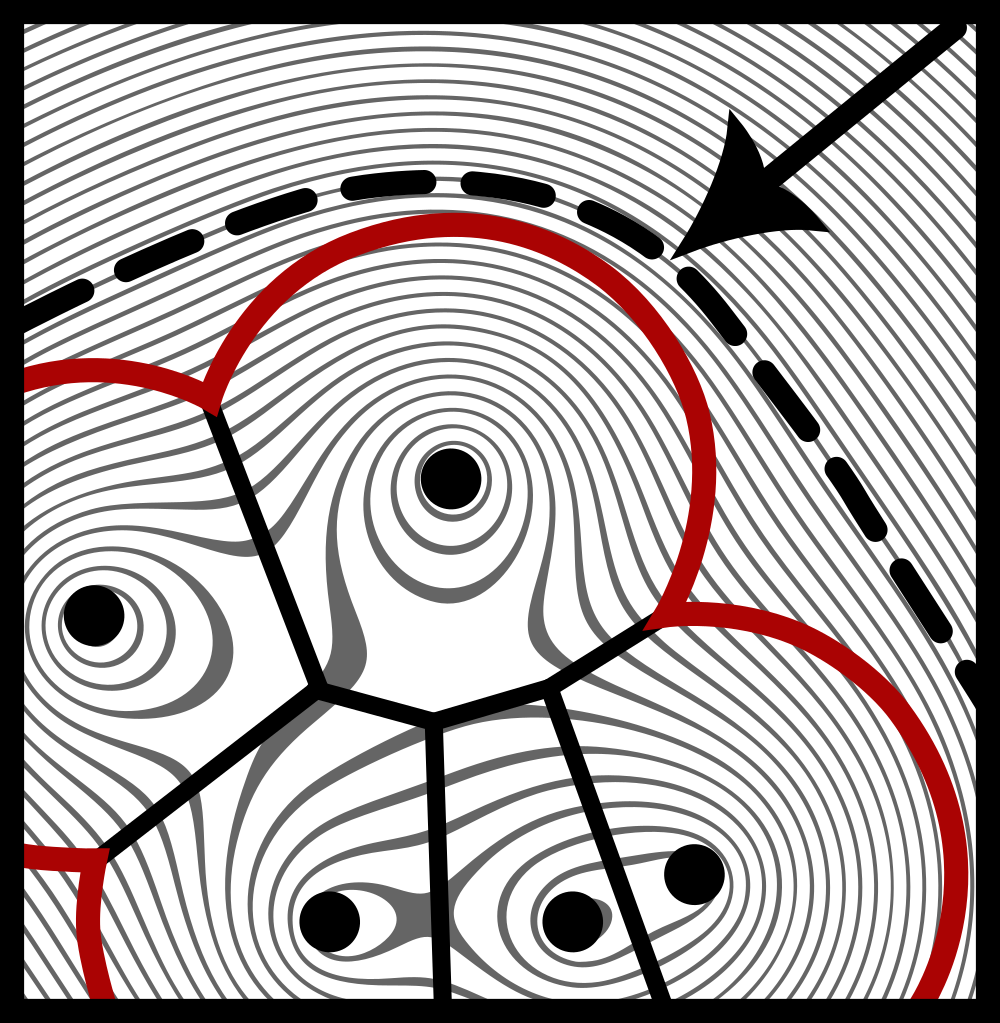}
  \end{center}
\end{wrapfigure}
\paragraph{Surface Smoothing} The free surface is characterized by the union of spherical patches, having discontinuous normals across the surface at the interface between two cells. To smoothen the geometry and obtain a more natural fluid rendering, we compute a smoothed surface by Sphere-Tracing (\Cref{fig:jet-10k}). The smooth surface is then characterized by the smooth union (see \textit{e.g.}~\cite{Dekkers2004, Quilez13}) between all spheres. In practice, we rely on a cubic polynomial in our implementation. To avoid the cost of considering every site of the simulation, we approximate the signed distance field by restricting the computation to the closest site and its Laguerre neighborhood. We are then able to perform the evaluation of the smoothed surface on the fly (see \textit{e.g.} \Cref{fig:render-shower}, \Cref{fig:physical-effects}, \Cref{fig:jet-10k}).

\begin{wrapfigure}{r}{0.275\columnwidth}
  \begin{center}
    \includegraphics[width=0.225\columnwidth]{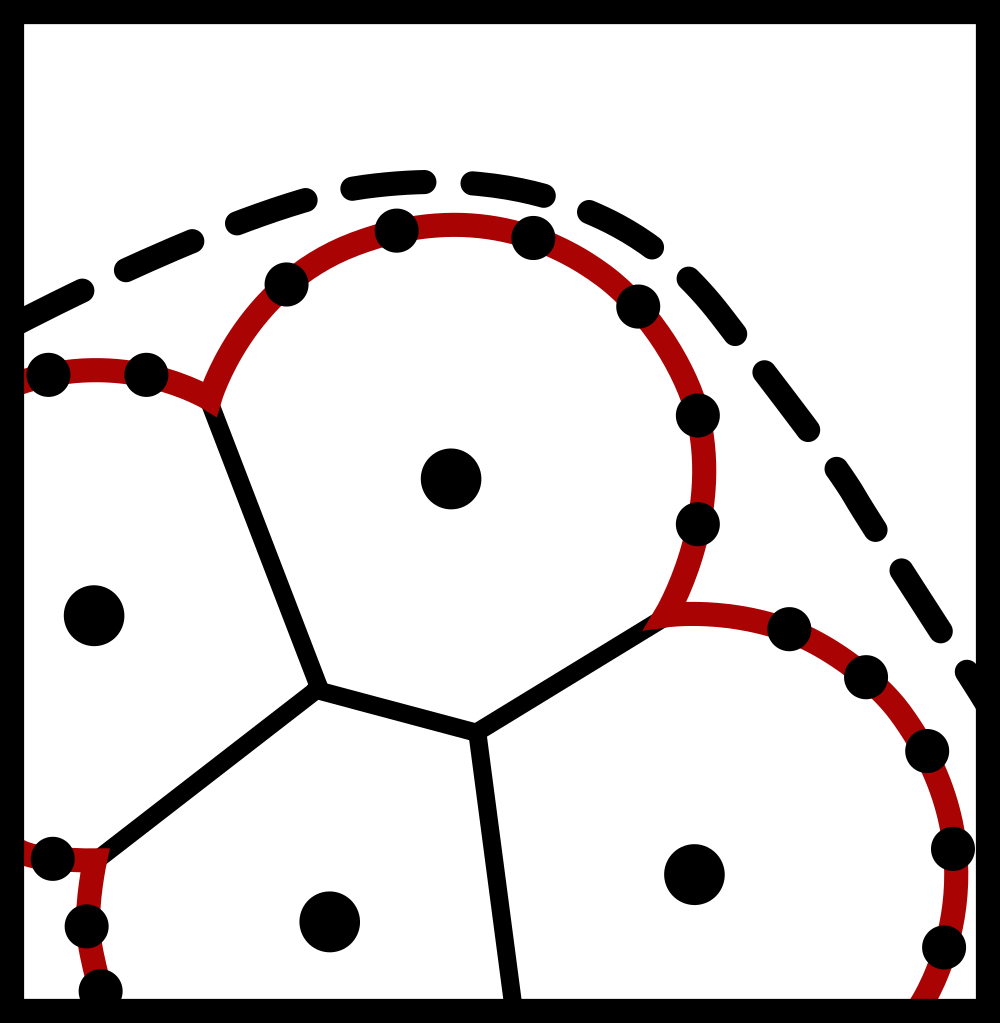}
  \end{center}
\end{wrapfigure}
\paragraph{Mesh Extraction} While the presented rendering process efficiently produces high quality images, it is not directly compatible with existing offline image production pipelines. Thankfully, our versatile representation can directly be used to produce a high-quality mesh. Similarly to previous works focusing on signed distance fields~\cite{Sellan2024}, we have to extract a smooth surface tangent to spheres. We then sample the surface of the fluid and perform a Poisson Surface Reconstruction~\cite{Kazhdan2006} producing a smooth surface mesh (see \textit{e.g.} \Cref{fig:teaser} and \Cref{fig:spring}). The so-extracted mesh can then be seamlessly integrated in existing rendering pipeline. We used this strategy in the video corresponding to \Cref{fig:teaser}, generated using the free software Blender~\cite{Blender}. The other videos were rendered with our method presented previously.

\section{Results}
\label{sec:results}

\newcommand{\configO}{Config. 1}
\newcommand{\configT}{Config. 2}

\begin{figure}[b]
    \centering
    \begin{tabular}{l@{\hspace{6pt}} r r r r}
        \toprule
        \multirow{2}{*}{\textbf{Status}} & \multirow{2}{*}{\textbf{Ours}} & \multicolumn{3}{c}{\textbf{\cite{Levy2022}}}\\\cmidrule(lr){3-5}
         &  & 42 & 162 & 320  \\
        \midrule
        Converged & 100 & 89 & 93 & 94  \\
        Not converged & 0 & 11 & 7 & 6  \\
        \bottomrule
    \end{tabular}\\[1em]
    \subfloat{\includegraphics[width=\columnwidth]{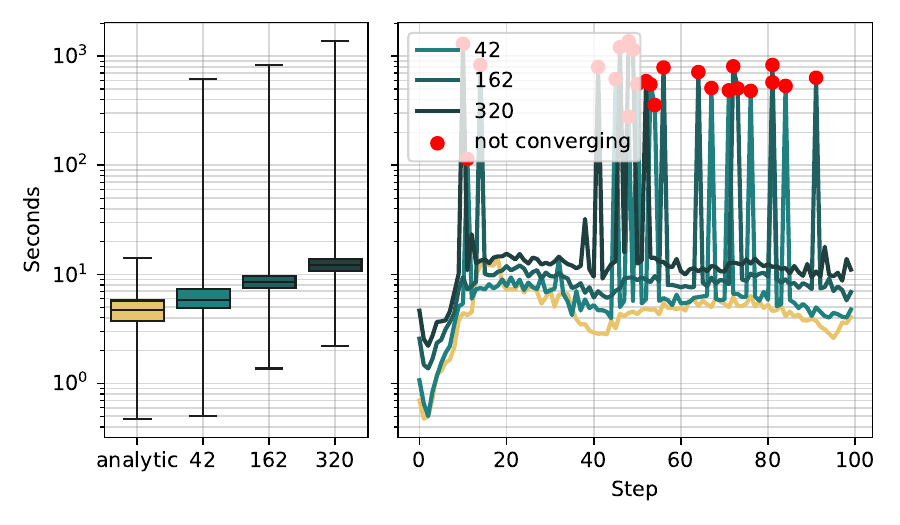}}
    \caption{Robustness benchmark (\Cref{fig:robustness-anim}) of our method compared to three discretization level (42, 162 and 320 vertices) in logarithmic scale (\configO). The implementation relying on discretization~\cite{Levy2022} fails to reach a $1\%$ error for the worst cell even after $100$ Newton iterations.}
    \label{fig:discrete-robustness}
\end{figure}

\begin{figure}[t]
    \subfloat{\includegraphics[width=.33\columnwidth]{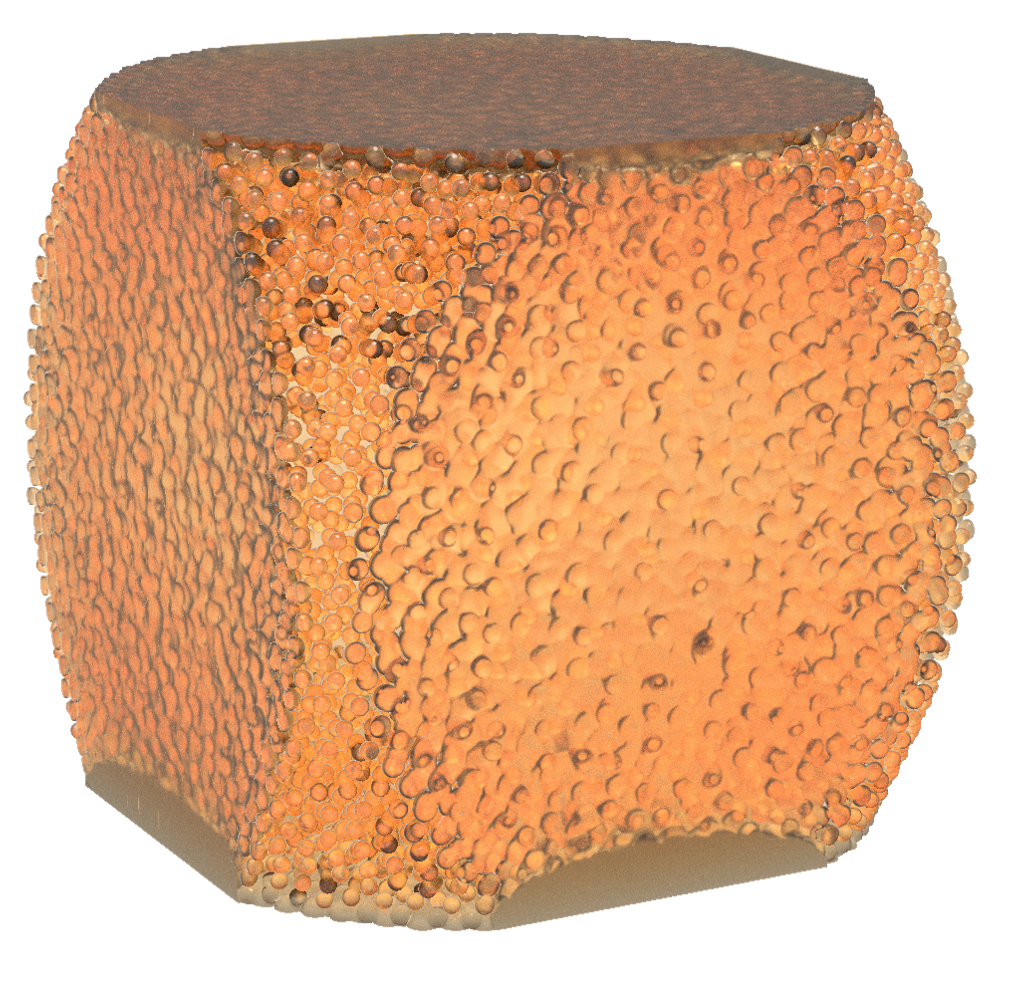}}\hfill
    \subfloat{\includegraphics[width=.33\columnwidth]{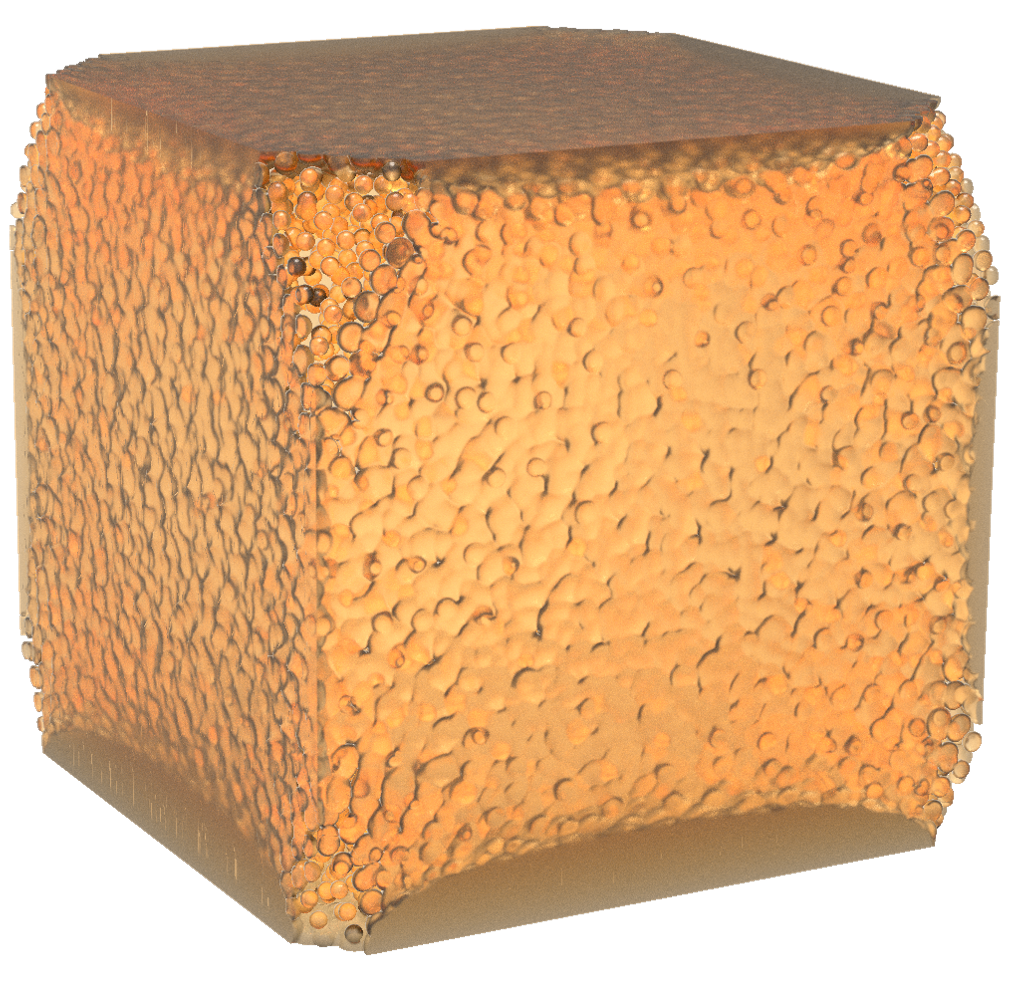}}\hfill
    \subfloat{\includegraphics[width=.33\columnwidth]{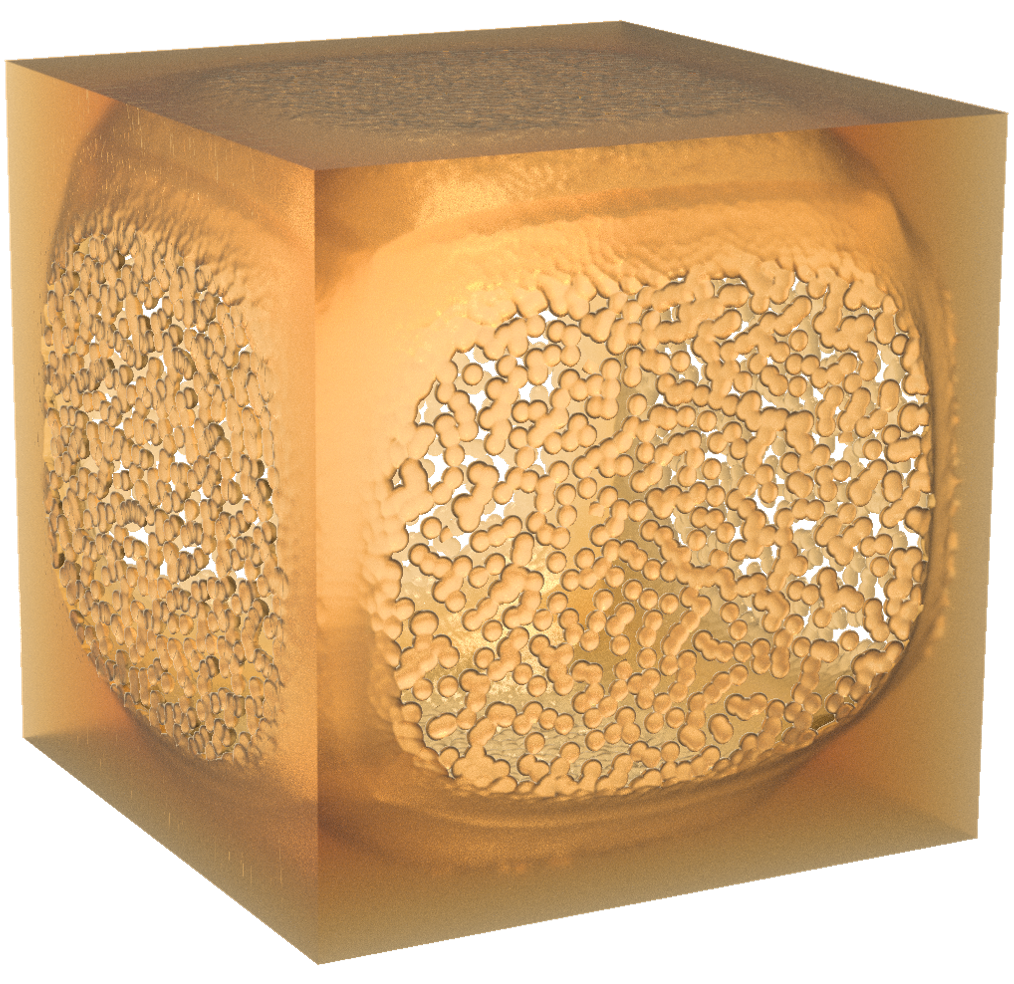}}\\[1em]
    \caption{Robustness benchmark (\configO) animation. Cells are initialized with explosive velocities from central points, leading to fluid boundary chocs. This strongly stresses the geometrical parts of the Newton solver.}
    \label{fig:robustness-anim}
\end{figure}
 
\begin{figure*}[t]
    \centering
    \includegraphics[width=\textwidth]{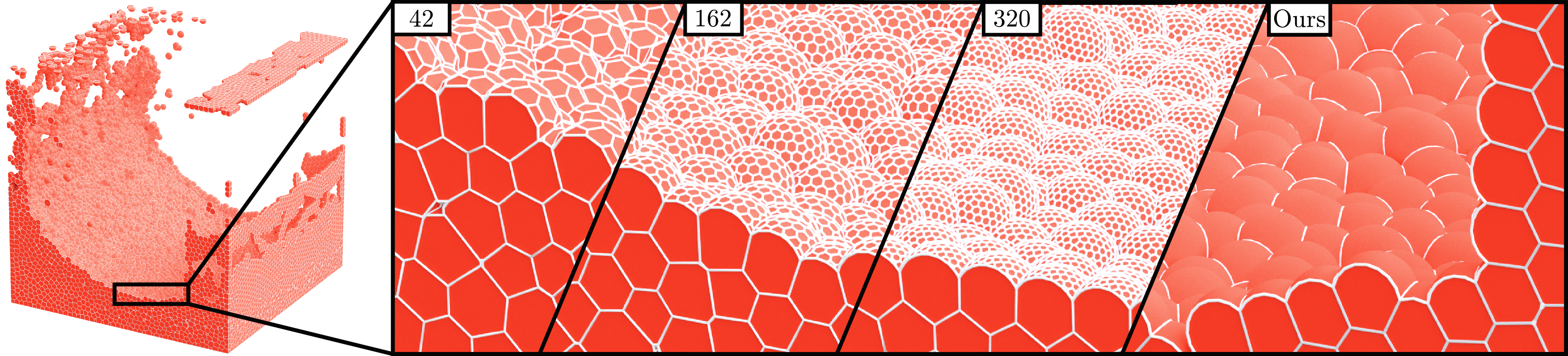}
    \caption{Illustration of the dam break simulation benchmark composed of $30 000$ cells, and the three discretization level used for comparison purposes.}
    \label{fig:discrete-performance-rendering}
\end{figure*}

We study the performances and robustness of our method both in terms of simulation and rendering. Our CPU implementation uses Geogram~\cite{geogram} to compute Laguerre diagrams while our GPU implementation is fully based on CUDA. Both implementations rely on a Conjugate Gradient solver and a Jacobi preconditioner. Benchmarks are given for two hardware configurations : an Apple MacBook Pro with an M1 Pro processor (\configO), and a workstation using a Intel i5 6600k and an NVIDIA RTX A4000 (\configT). In the following benchmarks, convergence of the Newton algorithm is reached when the worst cell has a volume error smaller than $1\%$ of its prescribed volume $\prescribedvolume$ and assumed as \emph{non-converging} when more than $100$ iterations have been performed without reaching this volume threshold. Note that this is much more demanding than previous works~\cite{Goes2015, Qu2022, Qu2023}. While this is more costly, it also allows performing detailed simulations with very few discretization points: as can be seen in the videos, rich and detailed fluid motions are obtained, even with a very small number of cells. \Cref{fig:teaser} shows a highly detailed crown splash, featuring accurate volume conservation of tiny droplets that split and merge.

\begin{figure}[!t]
    \centering
    \includegraphics[width=\columnwidth]{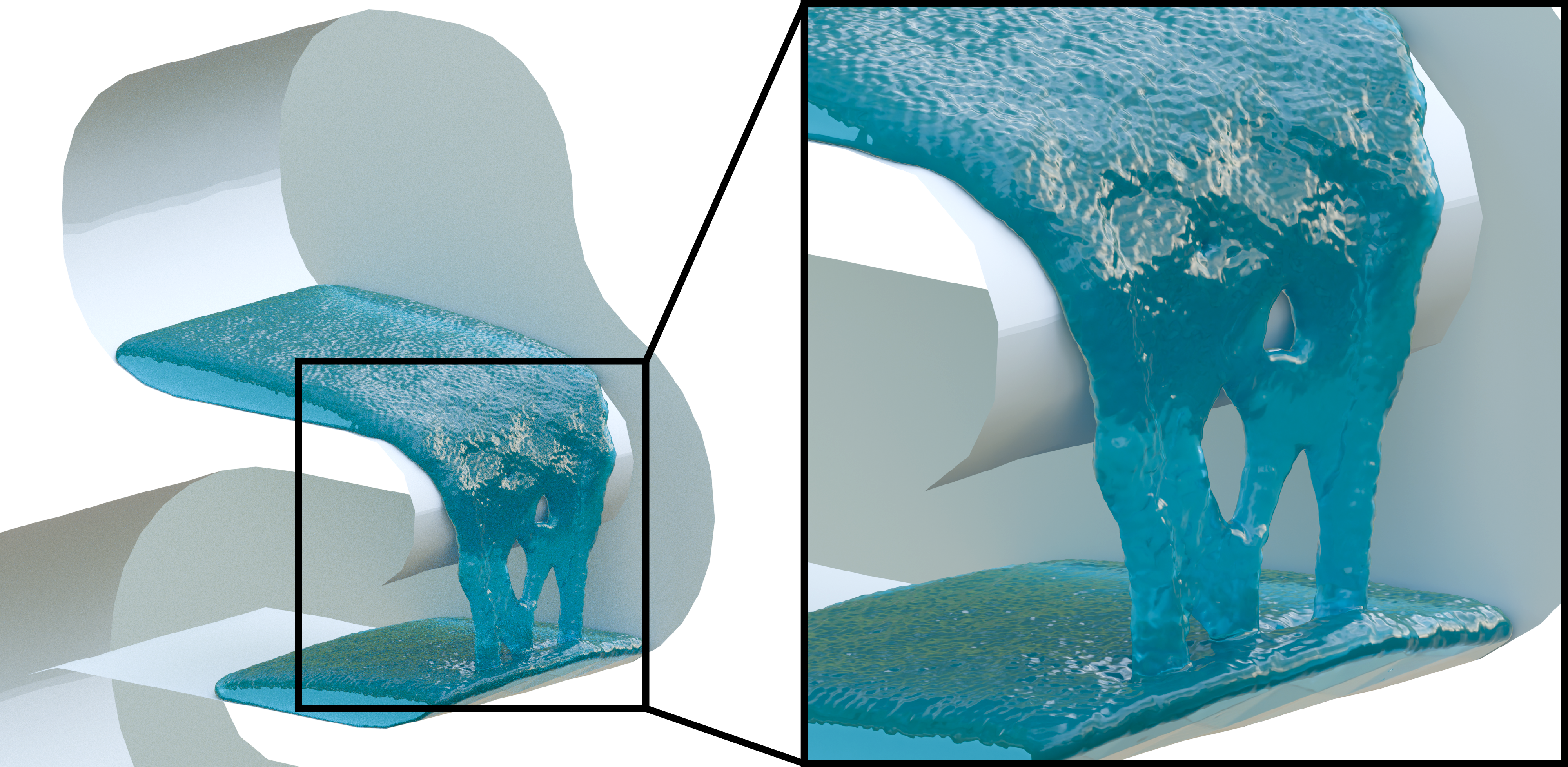}
    \caption{Illustration of a viscous fluid constrained in a polyhedral domain featuring complex fluid-boundary interactions. }
    \label{fig:spring}
\end{figure}

\begin{figure}[b]
    \centering
    \includegraphics[width=\columnwidth]{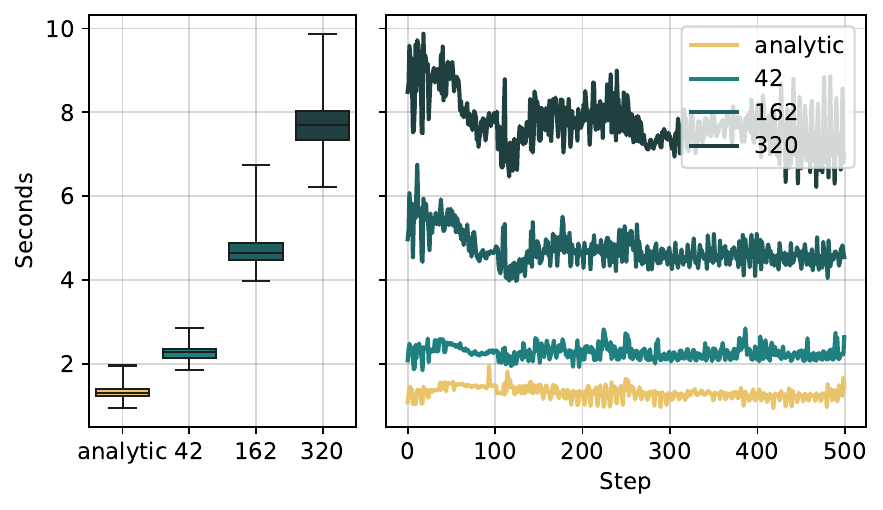}
    \caption{Performance benchmark (\configO) of our method compared to a discrete implementation~\cite{Levy2022} using three discretization levels (42, 162 and 320 vertices) on the dam break simulation illustrated \Cref{fig:discrete-performance-rendering}. Timing values include the complete simulation step and are given in seconds.}
    \label{fig:discrete-performance-data}
\end{figure}

\paragraph{Comparison with discrete implementations} We first compare our method with a discrete implementation~\cite{Levy2022} by relying on three discretization levels using 42, 162 and 320 vertices. They cover the trade-offs between precision and computational costs offered by this strategy. We first evaluate the robustness of our method in \Cref{fig:discrete-robustness} relying on a dedicated test case illustrated \Cref{fig:robustness-anim}. Fluid cells are initialized as a sphere with a high outward velocity, producing strong chocs against the domain boundary. The optimal transport problem in this case is numerically challenging because sites gather along domain boundary producing degenerate geometric configurations. This extreme case is seamlessly handled by our implementation and runs without any convergence error (an accuracy of $1\%$ volume error for the worst cell is reached at each time step). In contrast, the discrete implementation fails at multiple steps to minimize the cell volume error below the selected threshold. Furthermore, we notice that our method remains one of the fastest, even without considering these convergence issues. We further compare both methods performances in \Cref{fig:discrete-performance-data} using a simpler test case of a dam break composed of $30 000$ sites arranged as a cube falling due to gravity, illustrated in \Cref{fig:discrete-performance-rendering}. Considering $500$ simulation steps, our method remains the fastest, even compared to the coarser discretization level. This is a made possible by our analytic handling of fluid boundary that accurately computes the differential quantities. In contrast, discrete implementation must evaluate finely discretized mesh boundaries, involving high computational cost.

\begin{figure}[t]
    \centering
    \includegraphics[width=\columnwidth]{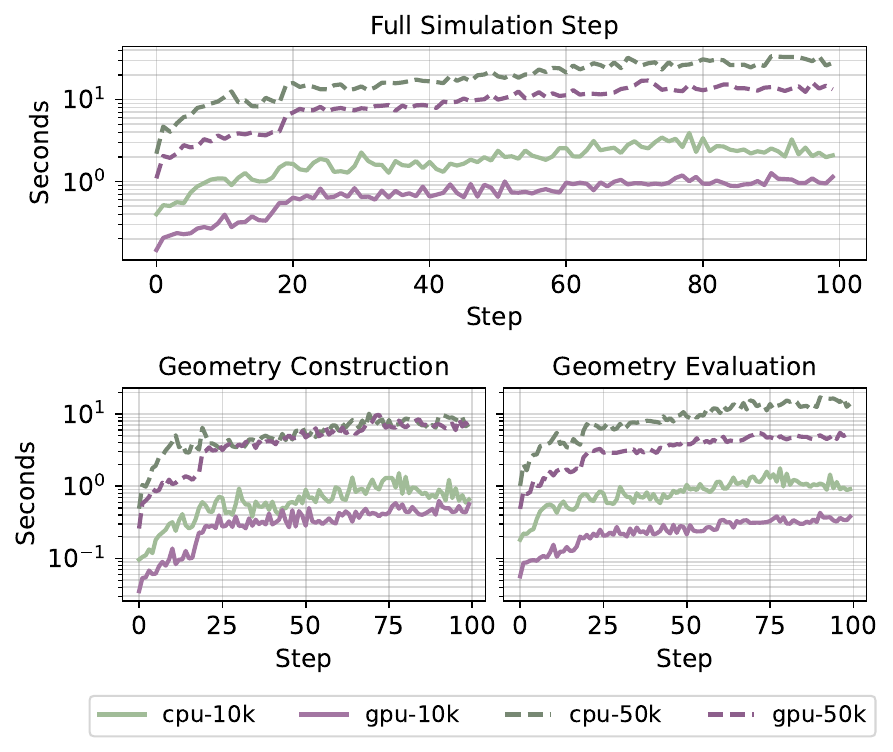}
    \caption{CPU (\configO) and GPU (\configT) performance benchmarks on $100$ iterations of the simulations presented in \Cref{fig:render-shower}.}
    \label{fig:cpu-vs-gpu}
\end{figure}

\paragraph{Performance} We evaluate in \Cref{tab:performance} the performance of our method based on the presented simulations. We observe that our GPU implementation remains steadily twice faster than our CPU implementation on the selected hardware, independently from the resolution of the simulation. This is notably due to the amount of computation involved in the evaluation of the differential quantities, strongly benefiting from a massively parallel implementation. We further evaluate both implementations in \Cref{fig:cpu-vs-gpu} on both animations presented \Cref{fig:render-shower}, starting with $10 000$ and $50 000$ sites. First, we confirm the gain offered by the GPU implementation considering the full computation time of a single step. Then, we compare the geometry construction timings, \textit{i.e.} the construction of the Laguerre diagram, and the geometry evaluation timings, \textit{i.e.} the analytic evaluation of the differential quantities. Considering the highly heterogeneous nature of Laguerre diagram construction, we first notice that it offers moderate speedup on the GPU, in line with their strengths and weaknesses for very irregular computation patterns, as involved in sparse linear systems. This very challenging task then remains interesting to avoid continuous memory transfers. Additionally, considering now the geometry evaluation, remaining steadily and significantly faster on the GPU, we observe sensible whole-clock gain, strongly benefiting from our highly parallel evaluation of the differential quantities. 

\begin{table}[b]
    \begin{tabular}{l@{\hspace{16pt}} rr rr}
    \toprule
    \multirow{2}{*}{\textbf{Sample}} & \multicolumn{2}{c}{\textbf{Surface}} & \multicolumn{2}{c}{\textbf{Volume}} \\ \cmidrule(lr){2-3} \cmidrule(lr){4-5} 
     & \configO & \configT & \configO & \configT \\
    \midrule
    \Cref{fig:teaser}     & 20.8 & 4.2 & 297.0 & 48.7 \\
    \Cref{fig:teapot}     & 11.1 & 1.6 & 54.9  & 9.8 \\
    \Cref{fig:bunny-drop} & 12.8 & 3.3 & 79.7  & 12.6 \\
    \bottomrule
    \end{tabular}\\[1em]
    \caption{Rendering benchmarks performs on the GPU with \configO{} and \configT{} on a surface rendering (opaque) and a volume rendering (a single evaluation of depth). Timings (in milliseconds) are computed on the first state of the given samples and as the means of $100$ random viewpoints in the sphere centered on the barycenter of the scene.}
    \label{tab:rendering-benchmark}
\end{table}

\paragraph{Rendering} We also evaluate in \Cref{tab:rendering-benchmark} the rendering performance of our method using the interactive renderer described in \Cref{sec:rendering}. Surface only rendering are performed in realtime using our two configurations, even on the largest test scene (\Cref{fig:teaser}). In contrast, volume rendering (only including a single evaluation of depth within the fluid), is more costly. As expected, rays traversing a larger scene will consider a larger number of cells to evaluate the depth, increasing computation time.

\paragraph{Physical effects} In addition to \Cref{fig:teaser}, we provide two animations showcasing viscosity and surface tension effects obtained with our implementation, illustrated in \Cref{fig:physical-effects}. The first animation is computed by assigning varying viscosity coefficients to both sides of the teapot and its inner liquid volume (the right side, in blue, is extremely viscous). The most viscous part of the teapot remains steady while the other part collapses on the inner liquid volume with a very low coefficient of viscosity, producing a splash. We also observe fine surface details which are preserved in the blue part during the full animation. The second animation is made by sampling a bunny shaped volume while disabling gravity. As surface tension is the main force in such case, the liquid slowly converges to a sphere. All the simulations presented here rely on the simple model described \Cref{sec:physics}, that can handle a very wide gamut of viscosities even within the same simulation. Finally, our accurate computation of the generalized Laguerre cells can be evaluated in arbitrary polyhedral domains, as demonstrated in~\Cref{fig:spring}, that features a viscous fluid simulation in a pipe. Similarly to what is done in \cite{Levy2022}, we compute the intersection of the Laguerre cells with the domain and the sphere (but without needing to discretize the spheres). The accurate representation of both the fluid free boundary and fluid-domain boundary contacts makes it possible to capture subtle effects.

\section{Conclusion \& Discussion}

Implementing free-surface fluid simulation through previous discretized partial optimal transport methods may encounter precision and stability issues, previously mitigated by increasing discretization precision at the expense of computation time. We have shown here that considering an \emph{infinite} number of discretization points leads to \emph{analytic} expressions. As a result, the so-revised algorithm is both more accurate and significantly faster. Besides improving the optimal transport-based fluid simulation~\cite{Goes2015, Levy2022} through a plug-and-play replacement of the solver, this could be also applied to other works relying on semi-discrete partial optimal transport, \textit{e.g.} for topological optimization~\cite{Dapogny2024}. These interesting use cases offer motivations for further improvements regarding not only accuracy, robustness, but also the presented physical model and rendering engine:
 
First, our GPU implementation provides interesting speed up, as evaluated. Due to its characteristics, it is, however, less robust than its CPU counterpart, benefiting from exact predicates during Laguerre diagram computation. In practice, we observe irregular computation performance and erroneous results in some complex cases. Additionally, depending on the configuration, its Laguerre diagram computation can be slower than existing multithreaded CPU implementations. Further developments for robust and fast Laguerre construction algorithms on GPU could strongly support such use cases.
   
Second, discretizing the free boundary of the fluid with spheres remains a first-order approximation with limited convergence behavior. It will be interesting (but very challenging) to devise an \emph{anisotropic} version of partial optimal transport, that approximates interfaces with ellipsoids instead of spheres, to better approximate zones of high curvature and thin sheets of fluids. 
    
Third, with our method, the physical parameters of the simulation remain difficult to configure. We sometimes observed some non-physical small bubbles appearing in the bulk of the fluid. While these artifacts can be artificially removed, we conjecture that this may be caused by an invalid combination of the incompressibility spring strength with the other forces. Theoretical studies of how to mutually tune the parameters and characterize their validity range could strongly improve the usability of such simulations.

Finally, high quality images can be produced by rendering volumetric shapes using restricted Laguerre diagrams. This has also been noticed in other contexts like in novel view synthesis~\cite{Govindarajan2025}. This motivates further developments, to avoid explicit conversion towards other representations like meshes. Additionally, while both proposed fluid surface smoothing procedures provide complementary advantages, on the fly smoothing through Sphere-Tracing is currently limited to the neighborhood of each cell, producing artifacts if the smoothing radius is too large, and the Poisson Surface Reconstruction cannot be used interactively. Further developments towards on-the-fly reconstruction of such representation could support various applications.\\

Implementation will be made available upon publication.

\bibliographystyle{ACM-Reference-Format}
\bibliography{bibliography}

\end{document}